\documentclass[a4paper,fleqn,usenatbib]{mnras}
\usepackage{CJK}
\usepackage{newtxtext,newtxmath}
\usepackage[T1]{fontenc}
\usepackage{ae,aecompl}
\usepackage[]{units}
\usepackage{graphicx}
\usepackage{amssymb}
\usepackage{amsmath}
\usepackage{threeparttable}
\usepackage{hyperref}
\usepackage{color}
\usepackage{hhline}
\usepackage{subfigure}
\usepackage{lineno}

\newcommand{\qeff}{Q_{\rm{eff}}}

\mathchardef\mhyphen="2D

\title[Generalized HOD]{Exploring the squeezed three-point galaxy correlation function with generalized halo occupation distribution models}

\author[S. Yuan, D. J. Eisenstein and L. H. Garrison]{
Sihan Yuan,$^{1}$\thanks{E-mail: sihan.yuan@cfa.harvard.edu}
Daniel J. Eisenstein,$^{1}$
and Lehman H. Garrison$^{1}$
\\
$^{1}$Harvard-Smithsonian Center for Astrophysics, 60 Garden St., Cambridge, MA 02138, USA
}

\date{Accepted XXX. Received YYY; in original form ZZZ}

\pubyear{2018}

\begin{document}
\label{firstpage}
\pagerange{\pageref{firstpage}--\pageref{lastpage}}
\maketitle

\begin{abstract} 

We present the GeneRalized ANd Differentiable Halo Occupation Distribution (GRAND-HOD) routine that generalizes the standard 5 parameter halo occupation distribution model (HOD) with various halo-scale physics and assembly bias. We describe the methodology of 4 different generalizations: satellite distribution generalization, velocity bias, closest approach distance generalization, and assembly bias. We showcase the signatures of these generalizations in the 2-point correlation function (2PCF) and the squeezed 3-point correlation function (squeezed 3PCF). We identify generalized HOD prescriptions that are nearly degenerate in the projected 2PCF and demonstrate that these degeneracies are broken in the redshift-space anisotropic 2PCF and the squeezed 3PCF. We also discuss the possibility of identifying degeneracies in the anisotropic 2PCF and further demonstrate the extra constraining power of the squeezed 3PCF on galaxy-halo connection models. 
We find that within our current HOD framework, the anisotropic 2PCF can predict the squeezed 3PCF better than its statistical error. This implies that a discordant squeezed 3PCF measurement could falsify the particular HOD model space. Alternatively, it is possible that further generalizations of the HOD model would open opportunities for the squeezed 3PCF to provide novel parameter measurements.
The GRAND-HOD Python package is publicly available at \url{https://github.com/SandyYuan/GRAND-HOD}.

\end{abstract}

\section{Introduction}
\label{sec:intro}

In the current cosmological paradigm, galaxies reside in dark matter haloes. Over the last two decades, the halo occupation distribution (HOD) has become a popular framework to model the galaxy-halo connection and constrain cosmological parameters \citep{1998Jing, 2000Seljak, 2000Peacock, 2001Scoccimarro, 2002Berlind, 2002Cooray, 2004Kravtsov, 2005Zheng, 2007Zheng, 2009Zheng}. Specifically, the standard HOD framework computes the probability distribution of finding $N_g$ galaxies in a given halo of mass $M$, $P(N_g|M)$. Based on galaxy formation models, galaxies in the HOD framework are further categorized into central and satellite galaxies, with different probability distributions and spatial distributions. For the rest of this paper, we simply refer to these as centrals and satellites. 

The HOD framework bypasses the need to directly model the complex physics of galaxy formation and provides a simple yet powerful tool to describe the galaxy-dark matter connection. The HOD framework has been successful in interpreting clustering data from cosmological surveys \citep[e.g.][]{2004Zehavi, 2004Zheng, 2005Tinker, 2005Zehavi, 2011Zehavi, 2013vdBosch, 2013Cacciato, 2013More}, constraining the stellar mass-to-halo mass relation \citep[e.g.][]{2003Yang, 2005Tinker, 2007vdBosch, 2007bZheng, 2011Wake, 2012Leauthaud}, and inferring dark matter halo masses \citep[e.g.][]{2001Bullock, 2004Porciani, 2008Wake}. 

Despite the many successes of the HOD approach, it makes several important assumptions that limit its predicative power. 
First of all, the satellite distribution is often assumed to track the spatial profile of the dark matter halo. This is not an explicit assumption of the HOD framework itself but is commonly assumed in HOD implementations. In additon, the centrals are also assumed to be situated at the minimum of the halo potential well. However, several recent studies suggest that the observed distribution of luminous satellites is significantly steeper than that of dark matter inside a halo \citep{2005vdBosch, 2010Watson, 2012Watson, 2015Piscionere}. This is to be expected since the satellites' dynamics are affected by additional baryonic processes such as pressure forces, radiative heating and cooling, and star formation. 
Another closely related assumption is that the satellites are bounded by some arbitrarily defined halo radius, such as the virial radius $R_{200c}$. However, several recent studies have suggested that the more natural halo boundary is the \textit{splashback radius}, the radius at which newly accreted matter reaches its first orbital apocenter after its initial turnaround \citep{2014Adhikari, 2015bMore, 2016More, 2017aDiemer, 2017bDiemer}. 

Many HOD implementations also assume that the central galaxy is at rest with respect to the center of mass of the dark matter halo and that the satellite galaxies track the velocity of their underlying dark matter particles. \citet{2015aGuo, 2015bGuo} theoretically modeled galaxy velocity bias in simulations, where the velocity of the galaxies are biased compared to the velocity of the underlying dark matter. They found that the standard HOD model incorporating velocity bias is able to reproduce the observed 2-point correlation function (2PCF) and interpret high-order statistics such as the 3-point correlation function (3PCF). Galaxy velocity bias has also been detected in hydro-dynamical simulations for host halos of different masses
\citep[e.g.][]{2003Yoshikawa, 2003Berlind, 2013Wu, 2017Ye}.

A fundamental assumption of the HOD framework itself is that halo occupation only depends on mass. However, it has recently become apparent that halo clustering and occupation also depend on properties such as halo formation time and halo concentration \citep[e.g.][]{2005Gao, 2006Wechsler, 2006Zhu, 2007Croton, 2007Gao, 2007Wetzel, 2007Zentner, 2008bDalal, 2008Li, 2011Lacerna, 2016Hearin, 2016Sunayama}. This effect is generically referred to as \textit{halo assembly bias}. 
Although HOD models that ignore assembly bias can successfully fit current galaxy clustering data, \citet{2014Zentner} demonstrated explicitly that ignoring assembly bias in HOD modeling may introduce significant systematic errors. \citet{2016Hearin} introduced a decorated HOD that incorporates assembly bias and found that the effect of assembly bias can be as large as $\sim 20\%$ in the linear regime for a SDSS-like sample. 

In this paper, we develop a generalized HOD implementation that incorporates a flexible satellite profile, central and satellite velocity bias, and a new implementation of assembly bias. We also seed the random number generator to suppress shot noise in our mock catalogs, thus ensuring that our mock statistics are differentiable with respect to HOD parameters. We showcase how HOD generalizations affect galaxy clustering statistics, such as the projected 2-point correlation function (2PCF), the anistropic redshift-space 2PCF, and the 3-point correlation function in the squeezed limit \citep[squeezed 3PCF, ][]{2017Yuan}. We further compare the constraining power of the 2PCF and the squeezed 3PCF by exploring 2PCF degeneracies in the generalized HOD parameter space and the potential of breaking these dependencies with the squeezed 3PCF. 

The remainder of this paper is organized as follows. In Section~\ref{sec:baselineHOD}, we review the baseline HOD and describe its implementation. In Section~\ref{sec:hod_decor}, we describe our methodology for 4 generalizations to the baseline HOD: the satellite distribution generalization, velocity bias, the closest approach distance generalization, and assembly bias. In Section~\ref{sec:simulation}, we describe the suite of N-body simulations used for this analysis. In Section~\ref{sec:signatures}, we showcase the signatures of the 4 HOD generalizations in the 2PCF and the squeezed 3PCF. In Section~\ref{sec:discussion}, we compare the constraining power of the 2PCF and the squeezed 3PCF on HOD models by exploring the possibilities of identifying HOD prescriptions that are degenerate in the 2PCF but distinguishable in the squeezed 3PCF. Finally, we draw some concluding remarks in Section~\ref{sec:conclusions}. 

\section{The baseline HOD}
\label{sec:baselineHOD}

\subsection{Formulation}
\label{subsec:formulation}

In this paper, we adopt the simple 5 parameter HOD described by \citet{2007bZheng, 2009Zheng} as our baseline HOD. This formulation of the HOD is theoretically motivated and has become standard over the last decade \citep{2004Kravtsov, 2005Zheng}. 

Specifically, the Z09 HOD gives the expected number of centrals and satellites as
\begin{align}
& \langle N_{\mathrm{cent}} \rangle = \frac{1}{2}\mathrm{erfc} \left[\frac{\ln(M_{\mathrm{cut}}/M)}{\sqrt{2}\sigma}\right], \nonumber \\
& \langle N_{\rm{sat}} \rangle = \left(\frac{M-\kappa M_{\rm{cut}}}{M_1}\right)^{\alpha},
\label{equ:hod}
\end{align}
where the parameters are fitted to the 2PCF of Luminous Red Galaxies (LRGs) in the Sloan Digital Sky Survey \citep[SDSS,][]{2000York, 2002Zehavi, 2003Abazajian, 2004Zehavi, 2005bEisenstein, 2005bZehavi, 2009Zheng, 2015Kwan}. $M_{\rm{cut}}$ sets the minimum halo mass to host a central galaxy. $\sigma$ is the scatter around the minimum mass and controls how fast the expected number of centrals reaches 1 from 0. $\kappa M_{\rm{cut}}$ sets the cut-off halo mass to host a satellite. $M_1$ sets the typical mass scale for a halo to host one satellite. $\alpha$ is the slope of the power-law and controls the number of satellites at high mass. The erfc(x) is the complimentary error function and is given by erfc($x$) = 1 -- erf($x$). Refer to Figure~1 in \citet{2017Yuan} for an illustrative plot of $\langle N_{\mathrm{cent}} \rangle$ and $\langle N_{\rm{sat}} \rangle$. 


\subsection{Implementation}
\label{subsec:baselinehod}

For each halo, Equation~\ref{equ:hod} dictates that the expected number of centrals is a fractional value within the range $0 \leq \langle N_{\mathrm{cent}} \rangle \leq 1$. Thus, $\langle N_{\mathrm{cent}} \rangle$ can be considered as the probability of the halo hosting a galaxy. We call the random number generator (RNG) to throw a uniform random number $\mathcal{R}_c$ between 0 and 1, and we assign a central galaxy to the center of mass of the halo if $\mathcal{R}_c < \langle N_{\mathrm{cent}} \rangle$. The central galaxy assumes the velocity of the center of mass of the halo. In redshift space, redshift-space distortion \citep[RSD,][]{1998Hamilton} on the central galaxies is implemented by modifying the LOS velocity of the central galaxies by the LOS velocity of the center of mass of the host halos. 

For the satellites, rather than assuming a fixed galaxy distribution profile such as an NFW profile \citep{1997Navarro}, we let the satellite distribution track the dark matter distribution by assigning satellites to dark matter particles and giving each particle of the halo equal chance to host a satellite. 
Specifically, for a halo with $N_p$ particles and $\langle N_{\rm{sat}} \rangle$ expected satellites, each particle has probability $p = \langle N_{\rm{sat}} \rangle/N_p$ of hosting a satellite. 
Then we use the RNG to assign a uniform random number between 0 and 1 to each particle of the halo $\mathcal{R}_{s,i}$ in a fixed order, where $i$ is the particle index. Particles with $\mathcal{R}_{s,i} < p$ have galaxies assigned to their positions. The satellites assume the velocities of their host particles. In redshift space, the RSD on the satellite galaxies are implemented by modifying the LOS position of the satellite galaxies by the LOS velocity of the host particles. 

There are many advantages to populating galaxies on a by-particle basis. First, we do not need to assume any profile for the galaxy distribution. Second, for our implementation $\langle N_{\mathrm{cent}} \rangle \ll N_p$, so each particle follows a Bernoulli distribution, we ensure that the number of galaxy distribution is Poissonian, $P(N_g|M) = \mathrm{Pois} (\langle N_g|M\rangle)$.



\section{Generalizations to the baseline HOD}
\label{sec:hod_decor}

In this section we present our generalized HOD formalism. Our implementation is currently suited for N-body cosmological simulations with ROCKSTAR \citep{2013Behroozi} outputs and access to a representative subsample of particles. The Python package is now available on github under the name GeneRalized ANd Differentiable HOD (\textsc{GRAND-HOD})\footnote{\url{https://github.com/SandyYuan/GRAND-HOD}}. While the package is currently adapted for the \textsc{AbacusCosmos} suite of simulations \citep{2017Garrison}, the ideas of our generalized HOD are broadly applicable. 

For the rest of this section, we first discuss our main considerations in our methodology and then describe in detail how we implemented our generalizations, specifically the satellite distribution generalization, velocity bias, dependence on closest approach, and assembly bias. 

\subsection{Basic considerations}
\label{subsec:considerations}
One of the fundamental goals in our HOD implementation is to make our mock statistics (such as the mock 2PCF) differentiable with respect to HOD parameters. Specifically, using a RNG to determine the positions and velocities of galaxies in a halo can lead to significant shot noise in the generated mocks even for tiny changes to the HOD parameters. This causes the derivatives of the mock statistics against HOD parameters to be noise-dominated. We use the word ``differentiable" to describe methods that suppress the shot noise and ensure stable derivatives in the mock statistics. 


In our implementation, before we generate a mock galaxy catalog, we seed the RNG with a fixed number and assign each particle a random number in a fixed order. This guarantees that each particle gets the same random number on each run regardless of parameter choices. 
For the generalized HODs, we also seed the RNG and take steps to ensure that each particle is still assigned the same random number regardless of generalizations and parameter choices. 

There are several applications in this analysis where we need to further increase the signal-to-noise in our clustering statistics. For such applications, we perform 16 runs of the HOD routine with 16 different initial seeds and average the results from the runs. We refer to this process as reseeding. Although computationally repetitive, it is much cheaper than re-running the full N-body simulations. Our schemes for suppressing shot noise and increasing signal-to-noise should be broadly applicable to future HOD work. 

Another consideration is the conservation of the expected number of galaxies with respect to HOD generalizations. Specifically, we want the expected number of galaxies per halo generated from the generalized HOD to be the same as that from the baseline HOD implementation, $\langle N_g|M\rangle_{\rm dec} = \langle N_g|M\rangle_{\rm std}$. 
Specifically, the probability of each particle in the halo to host a galaxy $p(j)$ has to satisfy 
\begin{equation}
\sum_{j = 0}^{N_p-1} p(j) = \langle N_g|M\rangle_{\rm std},
\end{equation}
where $j$ is the particle index within the halo, and $N_p$ is the number of particles in the halo. 
By conserving $\langle N_g|M\rangle$, we minimize the modifications needed to include a new generalization and automatically inherit the infrastructure that has already been developed for the baseline HOD. This idea of galaxy number conservation is also upheld in the assembly bias decoration introduced in \citet{2016Hearin}. 

\subsection{Satellite distribution}
\label{subsec:gen_sat_prof}

In our baseline algorithm, we assign all particles in a halo the same probability of hosting a satellite. This lets the satellite distribution profile track the halo profile. 
To deviate the satellite spatial distribution away from the halo profile, we implement the following procedure:
\begin{enumerate}
\item For each halo, we take the list of particles and assign a uniform random number between 0 and 1 to each particle. Denote the number by $\mathcal{R}_{s,i}$, where $i$ is particle index. 
\item We rank the particles by distance to the halo center in descending order, i.e. the outermost particle has rank 0 and the innermost particle has ranking $N_p - 1$. We denote the ranking of the $i$-th particle as $r_i$. 
\item Then we define the probability of a particle hosting a galaxy as a function of its rank. Specifically, the probability for the $i$-th particle with rank $r_i$ is given by
\begin{equation}
p_i = \bar{p}\left[1+s(1 - \frac{2r_i}{N_p-1})\right]\ \ \ \ i = 0, 1, 2, ..., N_p - 1.
\label{equ:pi_s}
\end{equation}
$s$ is the new satellite distribution parameter that modulates the shape of the satellite spatial distribution. 
\item We loop through all particles in the original (pre-ranking) order to populate galaxies. Just as in the baseline case, for each particle, if $\mathcal{R}_{s,i} < p_i$, then the particle hosts a galaxy, and vice versa. We use the original order for this step to preserve random number for each particle.
\end{enumerate}

The algorithm is designed so that when $s = 0$, $p_i = \bar{p}$, and we recover the baseline case where the satellite distribution tracks the halo profile. When $s > 0$, the particles that are further from halo center have a higher chance of hosting a galaxy and vice versa. Thus, $s > 0$ corresponds to when the satellite distribution is less concentrated than the dark matter distribution, and $s < 0$ corresponds to when the galaxy distribution is more concentrated. 


The formula also guarantees that $p(j)\geq 0$ when $|s|<1$. It is mathematically possible that $p(j) > 1$, but realistically such occurrences are extremely rare because $\bar{p}\ll 0.5$ in most cases. In fact, our test mocks show that the total number of galaxies changes by $\sim 0.01\%$ when $s = \pm 0.5$.

We also implement a second approach where we modulate the satellite distribution by varying the radius of the halo. Our baseline HOD implementation assumes that all galaxies are within $R_{200c}$. We can define a radius modulation factor $f$, where we only populate the innermost $1-f$ fraction of the particles with galaxies. Each particle now has a $p_{\rm{new}} = N_g/((1-f) N_p)$ chance of hosting a galaxy to preserve the expected number of galaxies per halo. We also assign random numbers to each particle before excluding outer particles to preserve the random numbers. A disadvantage of this approach is that we cannot easily expand the halo radius without rerunning the halo finder with a larger linking length. A comparison of the two approaches is presented in Section~\ref{subsec:sig_s} and Figure~\ref{fig:galprof_compare}.

\subsection{Velocity bias}
\label{subsec:gen_vel_bias}

Similar to the velocity bias implementation of \citet{2015aGuo, 2015cGuo}, we define two velocity bias parameters, $\alpha_c$ and $s_v$ for biasing the velocity of centrals and satellites, respectively. For the central, we first measure the velocity dispersion of the dark matter particles in the halo $\sigma_h$. Then we sample the LOS velocity of the central relative to the halo center from a Gaussian distribution:
\begin{equation}
f(v_c-v_h) = \frac{1}{\sqrt{2\pi}\sigma_c}\exp\left(-\frac{(v_c-v_h)^2}{2\sigma_c^2}\right),
\label{equ:alpha_c}
\end{equation}
where $\sigma_c = \alpha_c\sigma_h/\sqrt{3}$. The $\sqrt{3}$ is to project the 3D velocity dispersion of the halo $\sigma_h$ along the LOS. $v_h$ and $v_c$ are the velocities of the center of mass of the halo and the central along the LOS. Some studies use a Laplace distribution instead of a Gaussian distribution based on the observed velocities of brightest cluster galaxies (BCGs) in Abell clusters \citep{2014Lauer, 2015cGuo}. The statistics considered in the following sections of this paper only weakly depends on central velocity bias parameter $\alpha_c$, and thus the choice does not change the conclusions of this paper. 

Several previous studies \citep{2007Tinker, 2015aGuo, 2015cGuo} have implemented satellite velocity bias via simply boosting the satellite velocity relative to the underlying dark matter particle in the center of mass frame of the halo. A problem with such implementations is that with boosted velocities, the satellites no longer obey Newton's second law in the dark matter gravitational potential. In our implementation, we build a routine that avoids this problem by preferentially assigning satellites to particles with higher or lower velocities relative to the center of mass. 

In much the same way we implemented satellite distribution generalization, we take the particle list for each halo and assign a random number uniformly sampled between 0 and 1 to each particle. Then we rank the particles by the magnitude of their velocity relative to the halo center, with rank 0 corresponding to highest relative velocity. Then we set the probability of each particle to host a satellite as a function of its relative velocity ranking $r_i$:
 \begin{equation}
p_i = \bar{p}\left[1+s_v(1 - \frac{2r_i}{N_p-1})\right]\ \ \ \ i = 0, 1, 2, ..., N_p - 1, 
\label{equ:pi_sv}
\end{equation}
where $s_v$ is the satellite velocity bias parameter. Then we loop through all particles in the original (pre-ranking) order to populate galaxies.

With this implementation, we preserve the expected number of galaxies per halo, but we favor particles with higher or lower relative velocities to host satellites, effectively biasing the satellite velocity distribution relative to that of the dark matter. Our approach, however, still preserves Newtonian physics of the satellites, as each N-body particle is still on a valid orbit. 

\subsection{Closest approach to halo center}
\label{subsec:gen_perihelion}

The previous two generalizations are motivated by dependence on distance to halo center and relative velocity. A more physically motivated dependence that combines both radial position and relative velocity is the dependence on the distance to halo center at closest approach $r_{\rm{min}}$. Two phenomena towards the halo center that motivate $r_{\rm{min}}$ dependence are ram-pressure stripping and tidal destruction. 

To estimate $r_{\rm{min}}$, we assume a NFW halo profile and write out the equation of energy conservation and the equation of angular momentum conservation:
\begin{align}
& \frac{1}{2}v_t^2 + \frac{1}{2}v_r^2 + \Phi(r_0) = \frac{1}{2}v_m^2 + \Phi(r_{\rm{min}}), \label{equ:energy_conserv}\\
& v_t r_0  = v_m r_{\rm{min}}. \label{equ:momentum_conserv}
\end{align}
$v_t$ and $v_r$ are the current tangential and radial velocity components of the particle relative to the halo center. $r_0$ is the current distance to halo center. $v_m$ and $r_{\rm{min}}$ are the tangential velocity and radial position of the particle relative to halo center at closest-approach. $\Phi(r)$ gives the potential of an NFW profile
\begin{equation}
\Phi(r) = -\frac{\alpha}{r}\ln\left(1+\frac{r}{R_s}\right),
\label{equ:nfw_potential}
\end{equation}
where $\alpha = 4\pi G\rho_0 R_s^3$. $\rho_0$ and scale radius $R_s$ are given in the ROCKSTAR catalog. 

Combining Equation~\ref{equ:energy_conserv}, \ref{equ:momentum_conserv}, and \ref{equ:nfw_potential}, we obtain an iterative equation 
\begin{equation}
X^2 = \frac{v_t^2}{v_t^2+v_r^2+\frac{2\alpha}{r_0}\left(X^{-1}\ln\left(1+\frac{Xr_0}{R_s}\right) - \ln\left(1+\frac{r_0}{R_s}\right)\right)},
\label{equ:rm_iter}
\end{equation}
where $X = r_{\rm{min}}/r_0$. 
This equation is non-linear and does not have an analytic solution. We approximate the solution by running 10 iterations starting at $X = 1$ and making sure that the estimate has converged. 

After estimating $r_{\rm{min}}$ for all particles, we can incorporate the dependence on $r_{\rm{min}}$ in the same fashion as we did for satellite distribution and velocity bias. Specifically, we compute the probability to host a satellite $p_i$ as a function of $r_{\rm{min}}$ ranking, using the same formula as Equation~\ref{equ:pi_s} and Equation~\ref{equ:pi_sv}, but with the closest-approach modulation parameter $s_p$ in place of $s$ or $s_v$.

\subsection{Assembly bias}
\label{subsec:gen_assem_bias}

To introduce assembly bias into the baseline HOD, we incorporate halo concentration as a secondary parameter of the HOD. The motivation to use halo concentration is that it is known to be strongly correlated with the formation histories of dark matter haloes, with earlier forming haloes having higher concentrations at fixed halo mass \citep{2002Wechsler, 2003Zhao, 2006Wechsler, 2009Zhao, 2017Villarreal}. Specifically, the probability of finding $N_g$ galaxies in a halo now depends on both halo mass $M$ and concentration $c$, $P(N_g|M, c)$. 
Our procedure of assembly bias generalization is as follows:
\begin{enumerate}
  \item We first rank all the halos by halo mass from high to low, and call this list $[\textbf{halos}]^{M}$, where the superscript denotes that this list is ranked by halo mass $M$. We calculate the corresponding list of expected number of centrals and satellites of the ranked halos. Call these two lists $[N_{\rm{cent}}]^{M}$ and $[N_{\rm{sat}}]^{M}$. 
  \item Now we define a pseudo-mass of the halo, which is the halo mass modulated by the concentration, given by
  \begin{equation}
  \log M_{\rm{pseudo}} = \begin{cases} 
  \log M + A & \text{if}\ \  c > \bar{c}(M), \\
   \log M - A & \text{if}\ \ c < \bar{c}(M),
   \end{cases}
  \label{equ:Mpseudo}
  \end{equation}
  where $A$ is the assembly bias parameter which governs the strength of assembly bias in our model, and $\bar{c}(M)$ is the median concentration within a mass bin at mass $M$. We choose a step function for incorporating concentration into the pseudo-mass instead of a linear function to minimize the effect of concentration outliers. 
 \item Now we re-rank the list of halos $[\textbf{halos}]^{M}$ by pseudo-mass and call this new list $[\textbf{halos}]^{M_{\rm{pseudo}}}$. This means halos with higher-than-average concentration tend to move up in ranking and vice versa. Note that we do not re-rank the expected galaxy number lists,  $[N_{\rm{cent}}]^{M}$ and $[N_{\rm{sat}}]^{M}$. 
 \item Then, we map the expected number of galaxies from $[N_{\rm{cent}}]^{M}$ and $[N_{\rm{sat}}]^{M}$ onto the re-ranked list of halos $[\textbf{halos}]^{M_{\rm{pseudo}}}$ from the top down, with the top rank halo getting the highest expected number of galaxies and the bottom ranked getting the lowest. 
 \item Finally, once we have the expected number of centrals and satellites for each halo, we loop through the halos from the highest mass to the lowest mass and populate galaxies in each halo using the same by-particle scheme as described in previous sections. 
\end{enumerate}

Our routine bears qualitative similarities to the ``decorated HOD" proposed in \citet{2016Hearin}. The main difference is that we utilize a shuffling approach to naturally preserve the expected number of galaxies. The $A$ parameter sets the strength of assembly bias and will thus be referred to as the assembly bias parameter. Note that $A$ could also be different for centrals and satellites, thereby generating two pseudo-mass lists. However, for simplicity in this paper, we set $A_{\rm{cent}} = A_{\rm{sat}} = A$. 

\vspace{3mm}

To summarize, we have so far introduced 4 generalizations to the baseline HOD, with 5 new generalization parameters. They are satellite distribution parameter $s$, central velocity bias parameter $\alpha_c$, satellite velocity bias parameter $s_v$, closest-approach modulation parameter $s_p$, and assembly bias parameter $A$. Together with the 5 parameters of the baseline HOD, our generalized HOD framework is fully parametrized by 10 parameters. 

Our implementation also allows the different generalizations to be invoked simultaneously. The $s, s_v, s_p$ parameters are modifying $p_i$, the probability per particle to host a satellite. We simply multiply the probability modifiers from each of these 3 generalizations to obtain the overall probability per particle to host a satellite. The central velocity bias parameter $\alpha_c$ modulates the position of the central galaxy without affecting the satellite distribution. Thus, central velocity bias can be invoked simultaneously with the $s, s_v, s_p$ parameters. The assembly bias parameter $A$, changes the expected number of galaxies per halo without modifying the distribution of galaxies within each halo, whereas the other 4 parameters modulate the galaxy distributions within each halo. Thus, the assembly bias generalization can be invoked simultaneously with the other four generalizations. 

Our methodology of generalizations can also be easily customized to include further physically motivated dependencies. One interesting customization is to introduce halo mass dependence to our generalization parameters, i.e. setting $s, \alpha_c. s_v, s_p,$ and $A$ to be functions of halo mass instead of constants. Another customization is to modify the analytic form of the $p_i$ functions. So far we have set $p_i$ to be linearly dependent on the particle rank, but one could easily implement a more complicated function as long as it preserves the expected number of galaxies per halo.

\section{Simulations}
\label{sec:simulation}

We use the \textsc{AbacusCosmos} suite of emulator cosmological simulations generated by the fast and high-precision \textsc{Abacus} N-body code \citep[][Ferrer et al., in preparation; Metchnik $\&$ Pinto, in preparation]{2017Garrison, 2016Garrison}. Specifically we use a series of 16 cyclic boxes with Planck 2016 cosmology \citep{2016Planck} at redshift $z = 0.5$, where each box is of size 1100~$h^{-1}$~Mpc, and contains 1440$^3$ dark matter particles of mass $4\times 10^{10}$ $h^{-1}M_{\odot}$. The force softening length is 0.06~$h^{-1}$~Mpc. Dark matter halos are found and characterized using the ROCKSTAR \citep{2013Behroozi} halo finder. 

For assembly bias generalization, we also define the concentration parameter as
$c = r_{\rm{virial}}/r_s,$
where $r_{\rm{virial}}=R_{200c}$ is the virial radius of the halo. $r_s$ is the so-called Klypin scale radius \citep{2011Klypin}, which uses $v_{\rm{max}}$ and $M_{\rm{vir}}$ to compute the scale radius assuming an NFW profile. The Klypin scale radius is more robust than the standard NFW scale radius for small halos \citep{2013Behroozi}. As a sanity check, we compare the distribution of our concentration parameter and its correlation with halo mass with those of other N-body simulations \citep[e.g. Bolshoi and MultiDark,][]{2011Klypin, 2012Prada} and found them to be consistent. 

Nevertheless, our concentration measure is still noisy for halos with low numbers of particles. Thus, a moderate mass cut $M < 4\times 10^{12} h^{-1} M_{\odot}$ is applied on our halo sample to remove halos with less than 100 particles. This selection cut also helps in removing small halos whose profile is strongly affected by the force softening length, resulting in nonphysical halo profiles.

\section{Clustering signatures}
\label{sec:signatures}

\subsection{Clustering statistics}
\label{subsec:stats}

\begin{figure*}
\centering
 \hspace*{-0.4cm}
\includegraphics[width=7.2in]{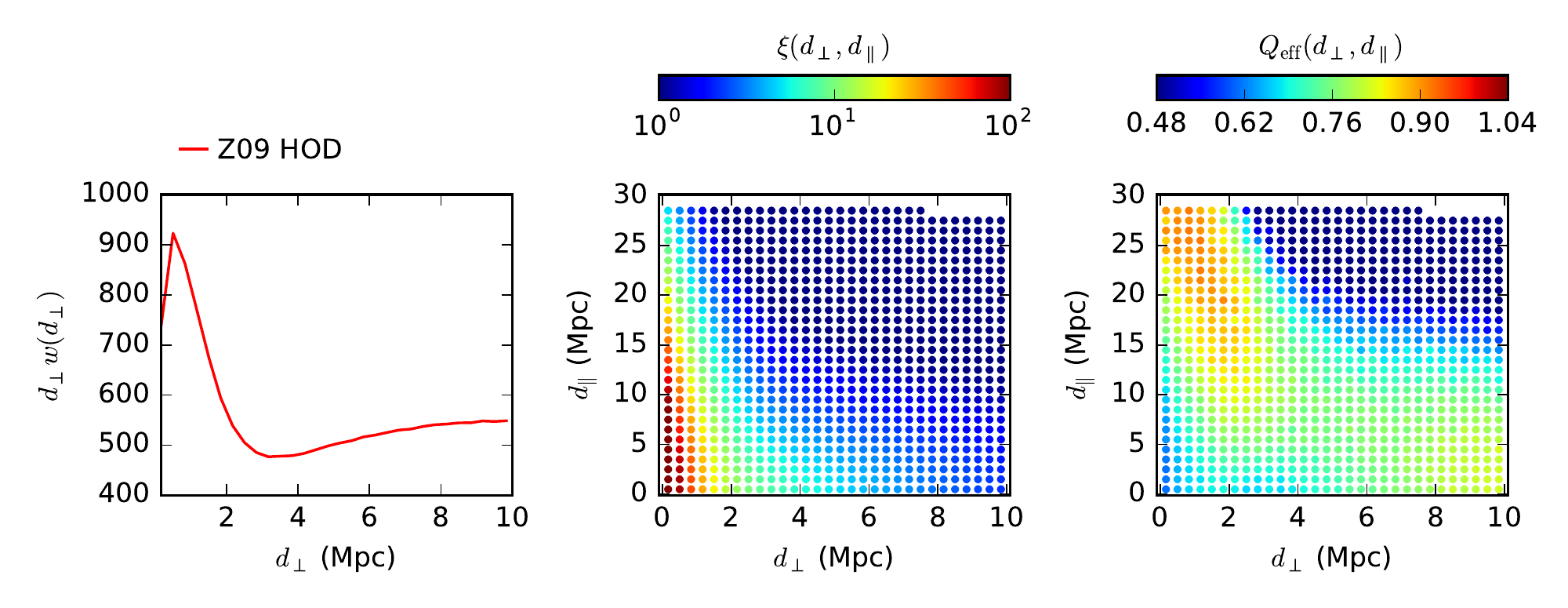}
\vspace{-0.6cm}
\caption{The projected 2PCF (left panel), anisotropic 2PCF (middle panel), and the squeezed 3PCF (right panel) corresponding to the baseline Z09 HOD parameters. }
\label{fig:triple_baseline}
\end{figure*}

In this section, we investigate the clustering signatures of the HOD generalizations we introduced in Section~\ref{sec:hod_decor}, using the \textsc{ABACUS} simulations. We study three clustering statistics: the projected 2PCF, the redshift-space anisotropic 2PCF, and the squeezed 3PCF. We briefly describe each of these statistics here. 

The 2PCF has been a critical tool in modern cosmology. In this paper, we model the galaxy-galaxy 2PCF with RSD, hereafter referred to as the anisotropic 2PCF. Given a mock galaxy catalog, we compute the anisotropic 2PCF using the SciPy kD-tree based pair-counting routine,
\begin{equation}
\xi(d_{\perp}, d_{\parallel}) = \frac{N_{\rm{mock}}(d_{\perp}, d_{\parallel})}{N_{\rm{exp}}(d_{\perp}, d_{\parallel})} - 1,
\label{equ:xi_def}
\end{equation}
where $d_{\parallel}$ and $d_{\perp}$ are the projected separation along the LOS and perpendicular to the LOS respectively. $N_{\rm{mock}}(d_{\perp}, d_{\parallel})$ is the number of galaxy pairs within each bin in $(d_{\perp}, d_{\parallel})$, and $N_{\rm{exp}}(d_{\perp}, d_{\parallel})$ is the expected number of pairs within the same bin but from a uniform distribution.

The projected 2PCF $w(d_{\perp})$ is defined as
\begin{equation}
w(d_{\perp}) = \int_{-d_{\parallel, \rm{max}}}^{d_{\parallel, \rm{max}}} \xi(d_{\perp}, d_{\parallel}) d(d_{\parallel}),
\label{equ:w_def}
\end{equation}
where $d_{\parallel, \rm{max}} = \sqrt{d_{p, \rm{max}}^2 - d_{\perp}^2}$ is the maximum separation along the LOS given $d_{\perp}$ and a maximum separation $d_{p,\rm{max}}$. For the rest of this paper, we choose $d_{p,\rm{max}} = 30$~Mpc. 

In \citet{2017Yuan}, we introduced the squeezed 3PCF as a new informative statistic to describe galaxy clustering. The squeezed limit refers to triangles that have one side much shorter than the other two sides, thus forming a configuration that can be thought of as a close pair with a distant third galaxy. The squeezed 3PCF effectively marginalizes over the position of the third galaxy, so we only study the dependence on the separation of the close pair. Thus, we compress the 5D general 3PCF into a more manageable 2D function. We showed in \citet{2017Yuan} that the squeezed 3PCF can be computed in $\mathcal{O}(N_g\log N_g)$ time for a galaxy field of $N_g$ galaxies. The squeezed 3PCF is denoted as $\qeff(d_{\perp}, d_{\parallel})$, and it can be interpreted as half of the average bias of the pairs within the separation bin. Thus, while the 2PCF describes the number of pairs, the squeezed 3PCF complements by describing the average bias and thus the average halo mass of the pairs. \citet{2017Yuan} also shows that the squeezed 3PCF is not a monotonic function of the pair separation and that it can be used as an additional test of the HOD framework. 

Figure~\ref{fig:triple_baseline} shows the projected 2PCF (left panel), anisotropic 2PCF (middle panel), and the squeezed 3PCF (right panel) corresponding to the baseline HOD parameters quoted in \citet{2009Zheng}. Specifically, the parameters are given as $M_{\rm{cut}} \approx 10^{13.35} M_{\odot}$, $M_1 \approx 10^{13.8} M_{\odot}$, $\sigma = 0.85$, $\alpha = 1$, and $\kappa = 1$. We refer to these parameter values as the Z09 values for the rest of the paper. The left panel shows that a large number of pairs are separated by $\sim 1$~Mpc, approximately the size of halos. The middle panel shows significant stretching in the anisotropic 2PCF along the LOS due to RSD. The right panel shows that the pairs separated by $\sim 1$~Mpc tend to be the most biased. This implies that these pairs tend to occupy the most massive halos. The large LOS separation of these pairs is due to the large RSD within large halos. 

\begin{figure*}
\centering
 \hspace*{-0.4cm}
\includegraphics[width=7.2in]{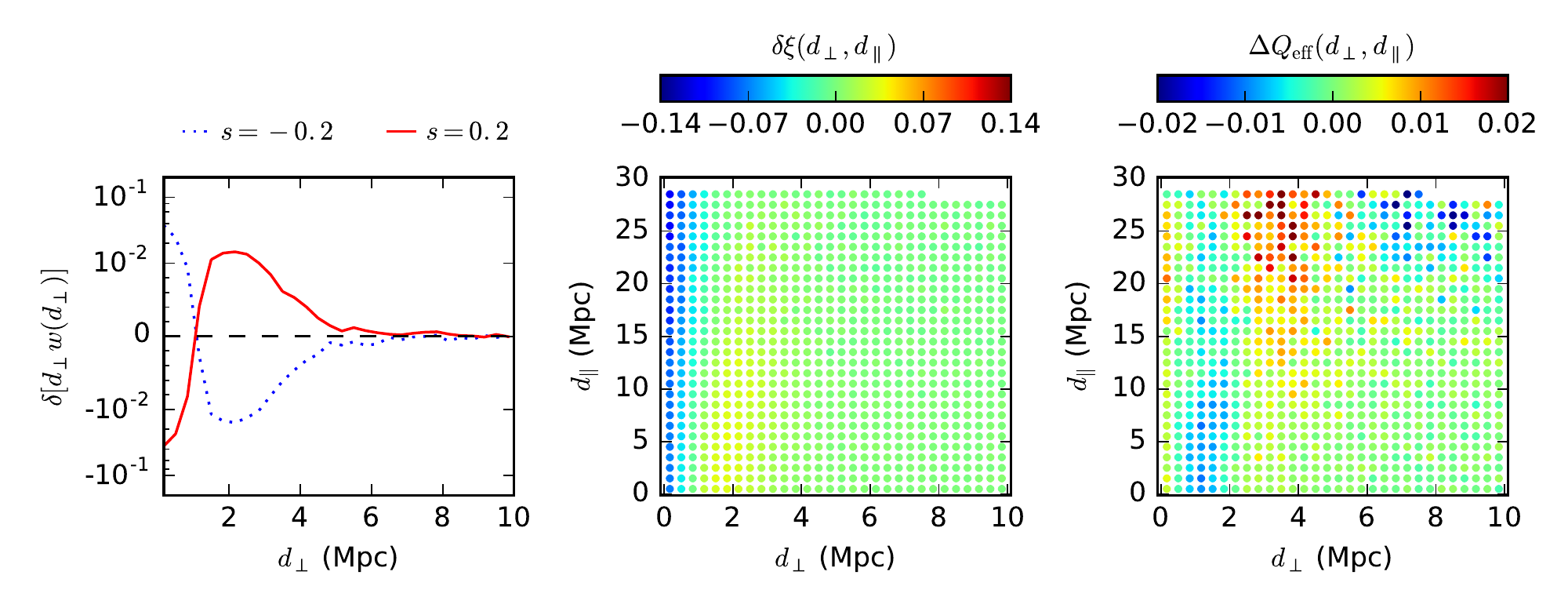}
\vspace{-0.6cm}
\caption{The perturbations to the projected 2PCF, anisotropic 2PCF, and the squeezed 3PCF due to modulating the satellite distribution parameter. Left panel: the fractional perturbation to the projected 2PCF for $s = 0.2$ (solid red) and $s = -0.2$ (dotted blue). For clarity, the y-axis is in linear scale from $-10^{-2}$ to $10^{-2}$ and in log scale beyond $10^{-2}$. Middle panel: the relative perturbation to the anisotropic 2PCF, where $\delta\xi = [\xi(s = 0.2)-\xi(s=-0.2)]/\xi(s=0)$. Right panel: the absolute perturbation to the squeezed 3PCF, where $\Delta \qeff = \qeff(s = 0.2)-\qeff(s=-0.2)$. We do not plot the fractional perturbation to $\qeff$ because $\qeff$ changes sign at high $d_\perp$ and $d_\parallel$, leading to a divergent fractional form.}
\label{fig:triple_s}
\end{figure*}

In the following subsections, we showcase the perturbations to these three statistics due to each HOD generalization. All the baseline parameters are held at Z09 values for now. All generalization parameters except for the one of interest in each subsection are set to 0. 

\subsection{Satellite distribution}
\label{subsec:sig_s}

Figure~\ref{fig:triple_s} shows the fractional perturbations to the projected 2PCF, anisotropic 2PCF, and the squeezed 3PCF due to varying the satellite distribution parameter $s$. The left panel shows the fractional perturbation to the projected 2PCF relative to the baseline HOD for $s = 0.2$ (solid red) and $s = -0.2$ (dotted blue). Specifically, for $s=0.2$, $\delta w = [w(s = 0.2)-w(s=0)]/w(s=0)$, and for $s = -0.2$, $\delta w = [w(s = -0.2)-w(s=0)]/w(s=0)$. We see that the two cases produce symmetric perturbations as expected. For $s = 0.2$, we are essentially moving galaxies away from halo centers, thus depleting the number of close pairs with $d_{\perp} < 1$~Mpc, and increasing the number of pairs with greater separation ($d_{\perp}\sim 2$~Mpc). The opposite happens for the $s = -0.2$ case. 

The middle panel shows the fractional perturbation to the anisotropic 2PCF, where $\delta\xi = [\xi(s = 0.2)-\xi(s=-0.2)]/\xi(s=0)$. We see that the perturbation is consistent with that shown for the projected 2PCF, with depletion at very small $d_{\perp}$ and enhancement at $d_{\perp}\sim 2$~Mpc. The signal is stretched out along $d_{\parallel}$ due to RSD. We also see that the satellite distribution parameter does not affect the 2PCF on larger transverse scales ($d_{\perp} > 6$~Mpc). This is because the distribution of satellites is a halo-scale phenomenon and its effects predominantly show up in the $d_{\perp}\sim 1$~Mpc regime of the 2PCF.

The right panel shows the absolute perturbation to the squeezed 3PCF, where $\Delta \qeff = \qeff(s = 0.2)-\qeff(s=-0.2)$.
 We do not plot the fractional perturbation for $\qeff$ due to the fact that it changes sign at high $d_\perp$ and $d_\parallel$ (refer to equation~32 of \citet{2017Yuan}). We do the same for all following plots of $\qeff$. The main feature is the decrease in $\qeff$ at $d_{\perp}\sim 1-2$~Mpc. This can be explained by the fact that, by moving galaxies towards the outskirts of halos, the $d_{\perp}\sim 1-2$~Mpc pairs that used to be dominated by pairs in large and massive halos are now being contaminated by pairs that lived in smaller and less massive halos. Thus, the average halo mass of these pairs decreases, leading to a decrease in the average pair bias. By moving galaxies outwards in the largest halos, we get a boost to the average pair bias and thus to $\qeff$ at $d_{\perp}\sim 3-4$~Mpc and $d_{\parallel}\sim 20-25$~Mpc. The large $d_{\parallel}$ typically results from peculiar velocities due to large RSD, and corresponds to a velocity dispersion of $\sim 1500$~km/s, consistent with $10^{15} M_{\odot}$ halos. 

\begin{figure}
\centering
\hspace{-1cm}
\includegraphics[width=3.4in]{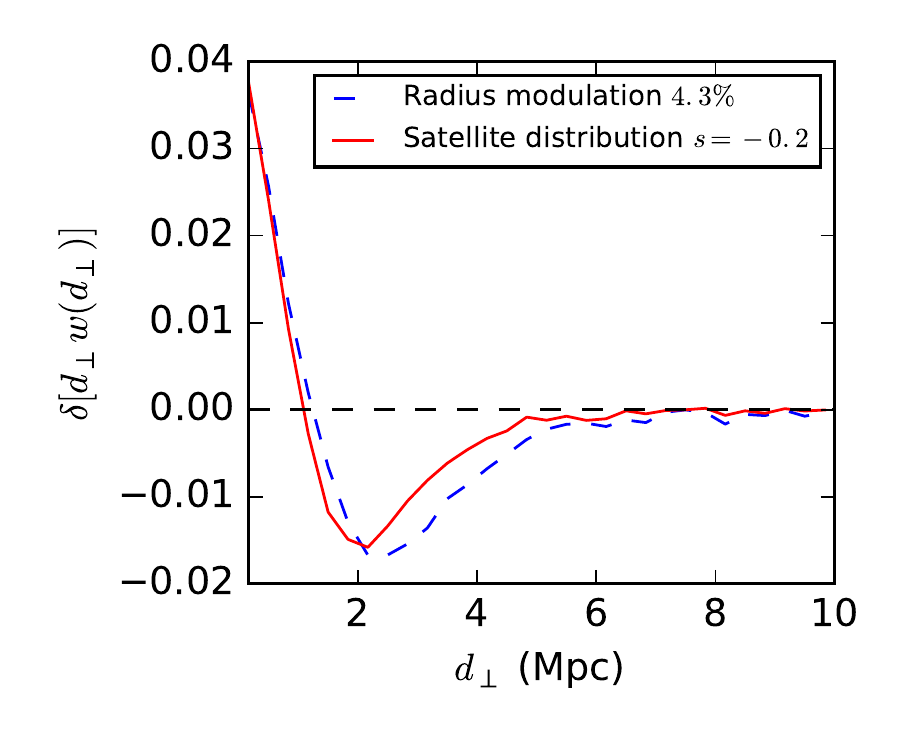}
\vspace{-0.6cm}
\caption{Comparing the effects of satellite distribution modulation and halo radius modulation. The red solid curve shows the effects due to modulating satellite distribution with $s = -0.2$, same as the blue dotted curve in the left panel of Figure~\ref{fig:triple_s}. The dashed blue curve shows the effects by shrinking halo radius by $4.3\%$, which is chosen to match the galaxy profile modulation curve. The two curves are qualitatively consistent.}
\label{fig:galprof_compare}
\end{figure}

In Section~\ref{subsec:gen_sat_prof}, we also considered an alternative way to modify satellite distribution, namely by shrinking halo radius. Figure~\ref{fig:galprof_compare} compares the fractional perturbation to the projected 2PCF due to satellite distribution modulation and halo radius modulation. The red solid curve reproduces the $s = -0.2$ case from the left panel Figure~\ref{fig:triple_s}. The blue dashed curve shows the halo radius modulation case where we shrink the halo radius by $4.3\%$. The shrinkage fraction is chosen to match the projected 2PCF signal of $s = -0.2$. We see that the two approaches to satellite distribution modulation produce qualitative similar signatures in the projected 2PCF. This is to be expected since the two approaches are both modulating the radial distribution of satellites, one with a linear probability distribution and the other with a step function distribution. We also test the anisotropic 2PCF and squeezed 3PCF signatures of shrinking halo radius and report that they are similar to those of satellite distribution modulation. We only include the satellite distribution generalization in the GRAND-HOD package as the two approaches are largely redundant. 

\subsection{Velocity bias}
\label{subsec:sig_sv}

\begin{figure*}
\centering
 \hspace*{-0.4cm}
\includegraphics[width=7.2in]{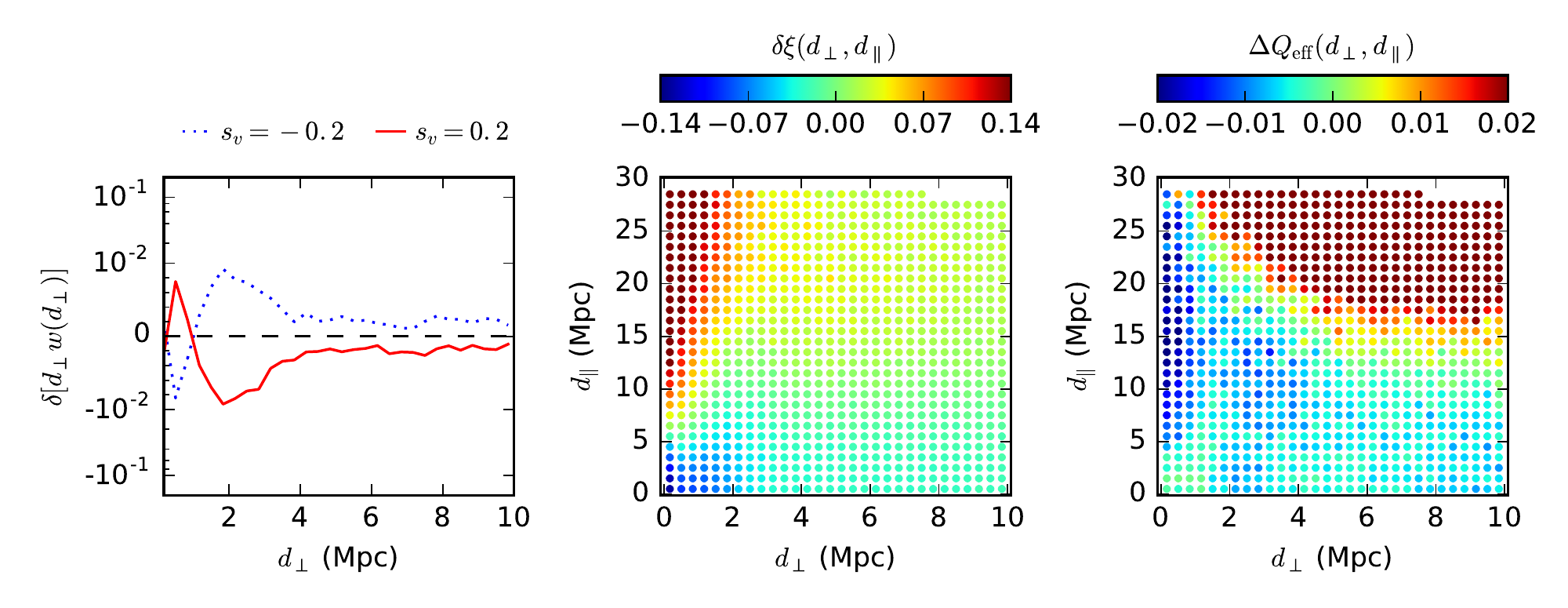}
\vspace{-0.6cm}
\caption{The perturbations to the projected 2PCF, anisotropic 2PCF, and the squeezed 3PCF due to modulating the satellite velocity bias parameter. Left panel: the fractional perturbation to the projected 2PCF for $s_v = 0.2$ (solid red) and $s_v = -0.2$ (dotted blue). For clarity, the y-axis is in linear scale from $-10^{-2}$ to $10^{-2}$ and in log scale beyond $10^{-2}$. Middle panel: the relative perturbation to the anisotropic 2PCF between the $s_v = 0.2$ case and the $s_v = -0.2$ case. Right panel: the absolute perturbation to the squeezed 3PCF between the $s_v = 0.2$ case and the $s_v = -0.2$ case.}
\label{fig:triple_sv}
\end{figure*}

Figure~\ref{fig:triple_sv} shows the fractional perturbations to the projected 2PCF, anisotropic 2PCF, and the squeezed 3PCF due to variations in the satellite velocity bias parameter $s_v$. We do not show the perturbations due to central velocity bias as it does not affect the projected 2PCF and creates negligible effects in the other statistics.

The left panel of Figure~\ref{fig:triple_sv} shows the relative perturbation to the projected 2PCF for the $s_v = 0.2$ case (solid red) and the $s_v = -0.2$ case (dotted blue). 
As we can see, although the velocity bias parameter only modulates the galaxy velocity distribution along the LOS, it produces a transverse signal in the projected 2PCF. This is because our implementation preserves Newtonian orbits of the satellite galaxies. Thus, a selection on particle relative velocity also entails a selection on radial position. The $s = 0.2$ case shows that by selecting particles of higher relative velocity, we are favoring pairs with transverse separation approximately $0.2$~Mpc $<d_{\perp}< 1$~Mpc.

The middle panel shows the relative change to the anisotropic 2PCF between the $s_v = 0.2$ case and the $s_v = -0.2$ case. As expected, we see significant structure along $d_{\parallel}$ direction. Specifically, since we are favoring particles with higher relative velocities to host satellites, we get a boost in the number of galaxy pairs at high $d_{\parallel}$ and strong RSD, and a decrease at small $d_{\parallel}$. It is also interesting that unlike the satellite distribution generalization, the velocity bias generalization also has percent level effects on the anisotropic 2PCF at $d_{\perp} > 5$~Mpc, beyond the 1-halo scale. 

The right panel shows the absolute change to the squeezed 3PCF, again for the $s_v = 0.2$ case and the $s_v = -0.2$ case. The blue trough at small $d_{\perp}$ is due to the fact that the small $d_{\perp}$ and large $d_{\parallel}$ pairs used to be dominated by pairs living at the bottom of the potential wells of the most massive halos, but are now contaminated with pairs from smaller halos with high RSD due to positive velocity bias. 

The increase in $\qeff$ at large $d_{\perp}$ and large $d_{\parallel}$ can be attributed to an interplay between high RSD and the anisotropic weighting function of the squeezed 3PCF. 
First it is important to note that our weighting function for the squeezed 3PCF (see Section~3.6 of \citet{2017Yuan}) favors the squeezed triangle to be point in the transverse direction. Thus, the $\qeff$ signal shows a strong pair bias in the at large $d_{\perp}$ and small $d_{\parallel}$ (see Figure~9 and 10 of \citet{2017Yuan}) where the three galaxies of the triangle are somewhat aligned, which tend to occur in strongly biased environments such as a transverse cosmic filament. There is a more detailed discussion of this phenomenon in section 4.1 of \citet{2017Yuan}. 
Without positive velocity bias, the galaxy pairs with large $d_{\perp}$ and large $d_{\parallel}$ tend to live in relatively uncorrelated halos. Thus the bias of these pairs relative to the galaxy field is low. Now due to positive velocity bias, the galaxies' LOS positions shift significantly. These relatively low-bias pairs now get contaminated with high-bias pairs that live on a single transverse filament (thus they used to have small $d_{\parallel}$) but now have large $d_{\parallel}$. This leads to higher pair bias on average at large $d_{\perp}$ and large $d_{\parallel}$. 

\subsection{Closest approach to halo center}
\label{subsec:sig_sp}

\begin{figure*}
\centering
 \hspace*{-0.4cm}
\includegraphics[width=7.2in]{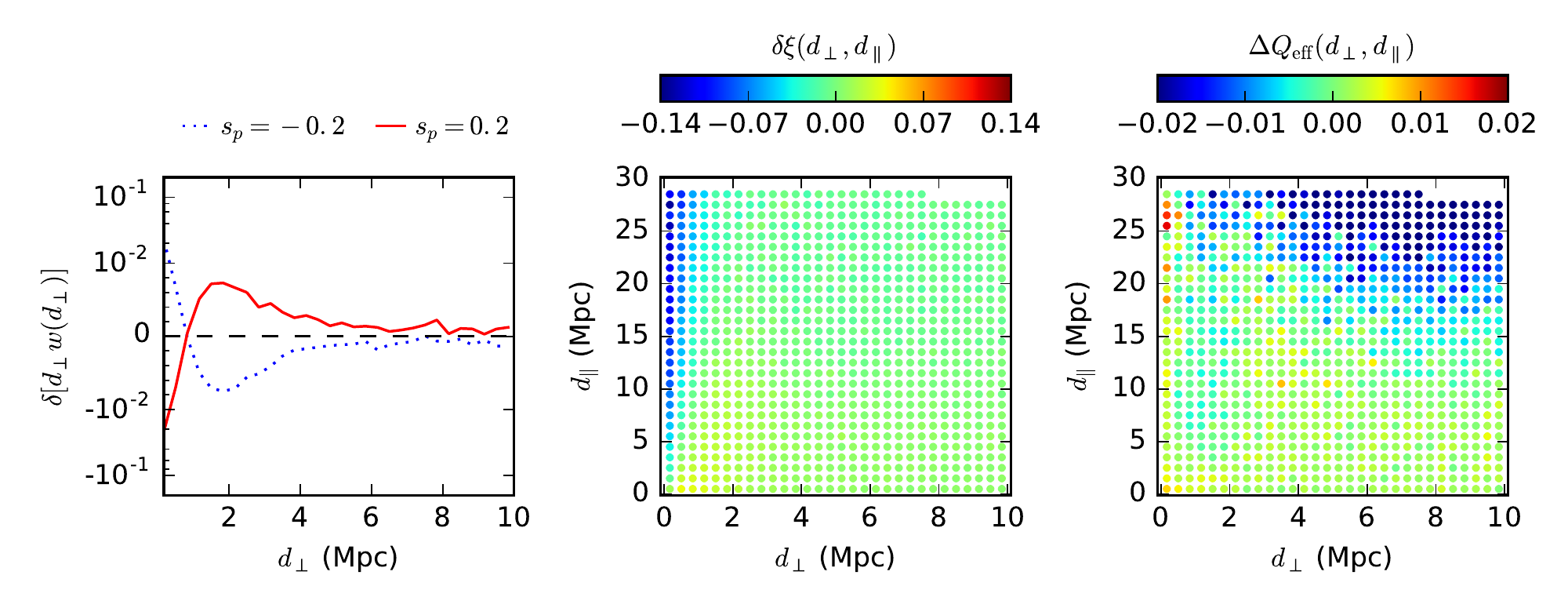}
\vspace{-0.6cm}
\caption{The perturbations to the projected 2PCF, anisotropic 2PCF, and the squeezed 3PCF due to modulating the the closest-approach parameter. Left panel: the fractional perturbation to the projected 2PCF for $s_p = 0.2$ (solid red) and $s_p = -0.2$ (dotted blue). The y-axis is in linear scale from $-10^{-2}$ to $10^{-2}$ and in log scale beyond $10^{-2}$. Middle panel: the relative perturbation to the anisotropic 2PCF between the $s_p = 0.2$ case and the $s_p = -0.2$ case. Right panel: the absolute perturbation to the squeezed 3PCF between the $s_p = 0.2$ case and the $s_p = -0.2$ case.}
\label{fig:triple_sp}
\end{figure*}

Figure~\ref{fig:triple_sp} shows the perturbations to the projected 2PCF, anisotropic 2PCF, and the squeezed 3PCF due to variations to the closest approach parameter $s_p$. 
The left panel shows the relative perturbation to the projected 2PCF for the $s_p = 0.2$ case (solid red) and the $s_p = -0.2$ case (dotted blue). The middle panel shows the relative change to the anisotropic 2PCF between the $s_p = 0.2$ case and the $s_p = -0.2$ case. The 2PCF signatures look similar to those of modulating the satellite distribution parameter in Figure~\ref{fig:triple_s}. This is to be expected since both generalizations select on particle positions, and the current radius of the particles are positively correlated with the closest approach radius. The closest approach generalization is conceived as a more physically motivated dependence compared to satellite distribution generalization that also incorporates velocity information. The $\qeff$ signature of the closest approach generalization is qualitatively similar to but weaker than that of satellite distribution generalization. The main features can be explained in the same fashion as for the satellite distribution generalization case. 

\subsection{Assembly bias}
\label{subsec:sig_A}

\begin{figure*}
\centering
 \hspace*{-0.4cm}
\includegraphics[width=7.2in]{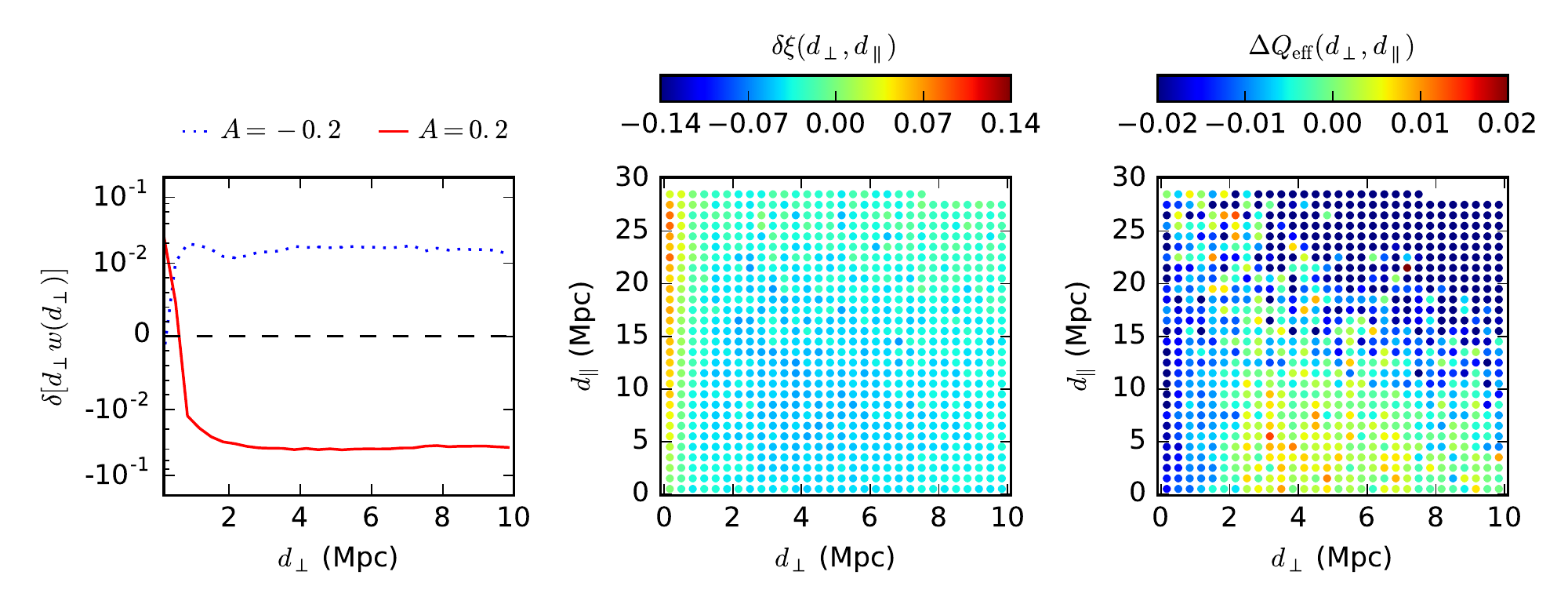}
\vspace{-0.6cm}
\caption{The perturbations to the projected 2PCF, anisotropic 2PCF, and the squeezed 3PCF due to assembly bias generalization. Left panel: the fractional perturbation to the projected 2PCF for $A = 0.2$ (solid red) and $A = -0.2$ (dotted blue). The y-axis is in linear scale from $-10^{-2}$ to $10^{-2}$ and in log scale beyond $10^{-2}$. Middle panel: the relative perturbation to the anisotropic 2PCF between the $A = 0.2$ case and the $A = -0.2$ case. Right panel: the absolute perturbation to the squeezed 3PCF between the $A = 0.2$ case and the $A = -0.2$ case.}
\label{fig:triple_A}
\end{figure*}

Figure~\ref{fig:triple_A} shows the perturbation to the projected 2PCF, anisotropic 2PCF, and the squeezed 3PCF due to assembly bias. The first thing to note is that the effect on the projected 2PCF is asymmetric between the $A = 0.2$ case and the $A = -0.2$ case. This is due to the asymmetric nature of our rank-based implementation of assembly bias. Because the halo mass function increases steeply towards the low mass end, $A < 0$ tends to change the ranking of a typical halo significantly more than $A > 0$. 

The shape of the perturbation to the projected 2PCF can be interpreted as the following. 
For $A = 0.2$, because we are moving galaxies from less concentrated halos into more concentrated halos, we get a boost in the number of the closest pairs ($d_{\perp} < 1$~Mpc) and a decrease in the number of large-separation pairs. For $A = -0.2$, we see that because we are moving galaxies away from concentrated halos and into more massive and puffier halos, we get a boost in number of slightly larger-separation pairs that track the typical size of halos ($d_{\perp} \sim 1$~Mpc). The same behavior is reflected in the anisotropic 2PCF in the middle panel.

Another interesting phenomenon to note here is that the assembly bias generalization also significantly modulates the 2PCF on the large scale, which underscores the importance of assembly bias in modeling the large-scale galaxy-galaxy clustering analysis and surveys. 

The right panel of Figure~\ref{fig:triple_A} shows the absolute change to the squeezed 3PCF between the $A = 0.2$ case and the $A = -0.2$ case. Since for positive assembly bias \citep{2006Wechsler, 2016Hearin}, we are moving galaxies into less massive and more concentrated halos, the close pairs on average now occupy less massive halos,. Thus, the average bias of close pairs decreases, leading to the drop in $\qeff$ at $d_{\perp} < 2$~Mpc. The larger-separation pairs at $d_{\perp} > 2$~Mpc now are more dominated by massive halos, thus the increase in average pair bias and $\qeff$ for those pairs.

\section{Investigation of degeneracies}
\label{sec:discussion}

In this section, we further test the constraining power of the squeezed 3PCF with the generalized HOD framework. We aim to identify combinations of HOD parameter changes that are degenerate or nearly degenerate in the 2PCF. Such degeneracies could then be broken with the squeezed 3PCF. This would imply that the 2PCF does not fully constrain the galaxy-halo connection model, and that the squeezed 3PCF offer extra constraints on the galaxy-halo model. 

Before we proceed, it is important to stress that the diagnostic power of the squeezed 3PCF does not rely on the discovery of degeneracies in the 2PCF. In \citet{2017Yuan}, we found that the projected 2PCF predicts the squeezed 3PCF to some precision better than the statistical error. This suggests that if we observationally match the projected 2PCF signal but find a discrepancy in the measured squeezed 3PCF larger than the statistical error, then we would have evidence to refute our HOD model.

\subsection{The 2PCF emulator}
\label{subsec:emulator}

To explore the HOD parameter space, it is overly expensive to repopulate the galaxies and recompute the 2PCF for each perturbation to the HOD parameters. Therefore we construct emulator models for both the projected 2PCF and the anisotropic 2PCF that approximates the two functions for small perturbations to the HOD parameters. The construction of such emulator for the projected 2PCF is detailed in Section~5.1 in \citet{2017Yuan}. We provide a short summary here. 

We have the HOD parameters as the emulator input and the binned projected 2PCF $w(d_{\perp})$ as the output. The projected 2PCF is parametrized by its values in $N_{\perp}$ equal bins along the $d_{\perp}$ axis from 0~Mpc to 10~Mpc, denoted as $w_1$, $w_2$,..., $w_{N_{\perp}}$. The emulated 2PCF $\hat{w_i}$ is formulated as a second-order Taylor expansion as a function of $\textbf{p}$,
\begin{align}
\hat{w_i}(\textbf{p}) = & w_i(\textbf{p}_0) + \sum_{j = 1}^n w_{i, p_j}(p_j - p_{0,j}) \nonumber \\
& + \frac{1}{2}\sum_{j = 1}^n\sum_{k = j}^n w_{i,p_j p_k}(p_j - p_{0,j})(p_k - p_{0,k}).
\label{equ:2pf_emulator}
\end{align}
$w_{i, p_j}$ denotes the the first derivative of each projected 2PCF bin $w_i$ against HOD parameter $p_j$.  $w_{i,p_j p_k}$ denotes the second derivative of $w_i$ against $p_j$ and $p_k$. $p_{0,j}$ refers to the Z09 value of $p_j$. $n$ denotes the number of HOD parameters. 
For our generalized HOD, we include up to $n = 9$ parameters in our emulator. We omit the parameter $\kappa$ since it does not significantly perturb the 2PCF or the squeezed 3PCF. 
One thing to note is that our emulator only performs well for small perturbations to the HOD parameters. For the rest of our analysis, we limit ourselves to HOD parameter perturbations $< 10\%$. 

The anisotropic 2PCF emulator is constructed in a similar fashion except for it outputs the anisotropic 2PCF, which is parametrized by its values in uniform 2D bins in the $(d_{\perp}, d_{\parallel})$ plane. Each value is again formulated as a second-order Taylor expansion as a function of $n$ generalized HOD parameters. 

\subsection{Model comparison}
\label{subsec:model_compare}

When searching for degeneracies, we need a means of quantifying the difference between the statistical predictions of different HOD parameter choices.
To illustrate our approach, we discuss the comparison of two different projected 2PCFs. The comparison of the anisotropic 2PCF and the squeezed 3PCF follows the same idea. 

We denote the difference between two projected 2PCFs with a length-$N_{\perp}$ array, $[\Delta(d_{\perp}w_i)]$, where $i = 1,2,3,..., N_{\perp}$. $d_{\perp}$ is again pair separation along the direction transverse to the LOS. We define 
\begin{equation}
\chi^2(d_{\perp}w) = [\Delta(d_{\perp}w_i)][C(d_{\perp}w_i, d_{\perp}w_j)]^{-1}[\Delta(d_{\perp}w_j)]^T,
\label{equ:chi2}
\end{equation}
where $[C(d_{\perp}w_i, d_{\perp}w_j)]$ denotes the $N_{\perp}\times N_{\perp}$ covariance matrix. To compute the covariance matrix, we sub-divide each of the sixteen $1100 h^{-1}$~Mpc boxes into 125 equal sub-volumes, yielding a total of 2000 sub-volumes and a total volume of $21.3 h^{-3}$~Gpc$^3$. We calculate the covariance matrix from the dispersion among the 2000 sub-volumes. The degree of sub-division is chosen to have a significant number of sub-volumes, yet ensuring that each sub-volume is large enough that the dispersion among the sub-volumes is not dominated by sample variance. 

For the anisotropic 2PCF and the squeezed 3PCF, since they are both represented by 2D arrays in the $(d_\perp, d_\parallel)$ plane, we flatten them into 1D arrays of length $N_\perp\times N_\parallel$. Then we utilize the same $\chi^2$ routine as described in the previous paragraph to define $\chi^2(\xi)$ and $\chi^2(\qeff)$. The covariance matrices for the anisotropic 2PCF and the squeezed 3PCF are of size $(N_\perp\times N_\parallel)^2$, and we evaluate each element from the dispersion among the 2000 simulation sub-volumes. 

It is important that we select the right binning for our statistics when computing the covariance matrices. On the one hand, in order to guarantee a reasonable signal-to-noise in the covariance matrix, we need to ensure that the total number of bins is much less than the number of sub-volumes. On the other hand, we need enough bins to capture all the important features in our statistics. 
For the projected 2PCF, we use $N_\perp = 30$ equal bins along the $d_{\perp}$ axis. For the anisotropic 2PCF and the squeezed 3PCF, we see that both statistics have rich structures along the $d_\perp$ direction (Figure~\ref{fig:triple_s}-\ref{fig:triple_A}). Thus, we adopt a relatively fine $N_\perp = 15$ equal bins along the perpendicular direction between 0-10~Mpc and a coarser $N_\parallel = 10$ equal bins along the LOS direction between 0-30~Mpc. This scheme yields a total of 150 bins for a size $150\times 150$ covariance matrix. 

Note that while we use the $15\times 10$ binning to compute the covariance matrices and the $\chi^2$ of the anisotropic 2PCF and the squeezed 3PCF, we will continue to plot these two functions in $30\times 30$ bins for the sake of visualization. 

\subsection{Degeneracies in the projected 2PCF}
\label{subsec:discuss_proj}
 
In this section, we aim to identify prescriptions of HOD parameters that are nearly degenerate in projected 2PCF, utilizing the emulator and $\chi^2$ routine we have developed. By testing various combinations of generalized and baseline HOD paraemters, we find that the satellite distribution parameter in addition to the baseline HOD parameters generates a near degeneracy in the projected 2PCF. 

Specifically, our emulator for this case includes the 4 baseline HOD parameters (omitting $\kappa$) and the satellite distribution parameter $s$ for a total of $n = 5$ input parameters. To construct the emulator, there are 5 first derivatives $w_{i, p_j}$ and 15 second derivatives $w_{i,p_j p_k}$ for each bin. We solve for the 20 derivatives by computing the projected 2PCF from the mock catalogs of 20 different HOD prescriptions. Again we seed the RNG and hence preserve the random numbers for each particle to obtain stable derivatives. 

To identify degeneracies in the projected 2PCF, we start with $s = \pm 0.2$ respectively while the other 4 baseline HOD parameter start at their Z09 values. We refer to these two starting HOD prescriptions as the ``unsuppressed" HODs. 
Then while holding the $s$ at $\pm 0.2$, we perturb the other 4 HOD parameters to minimize the $\chi^2(d_{\perp}w)$, defined by Equation~\ref{equ:chi2}. Specifically, the $\chi^2(d_{\perp}w)$ is calculated against the projected 2PCF of the baseline HOD.

The $\chi^2(d_{\perp}w)$ is minimized with the Powell algorithm \citep{1964Powell} and confirmed with the Nelder-Mead algorithm. We repeat the minimization starting at evenly spaced initial points around the origin in the 4D HOD parameter space. We confirm that the results converge regardless of initial conditions. The best-fit HOD prescriptions that minimize the $\chi^2$ for $s = 0.2$ and $s = -0.2$ cases are given in Table~\ref{tab:besthod_galprof} and referred to as the ``suppressed" HODs. We also test the performance of the emulator by comparing the emulated 2PCFs with the projected 2PCFs computed from mocks, and we find the deviations between the two relative to the baseline 2PCF is $< 0.4\%$ from peak to peak, with a rms of $0.16\%$. 

\begin{table}
\centering
\scalebox{1}{
\begin{tabular}{ c | c c c c}
\hhline {=====}
$s$ & $\log_{10} M_{\rm{cut}}$ & $ \log_{10} M_1$ & $ \sigma$  & $\alpha$ \\ 
\hline
0.2  & 13.35--0.0057 & 13.8--0.025 & 0.85--0.032 & 1--0.046 \\ 

--0.2 & 13.35+0.0017 & 13.8+0.033 & 0.85+0.011 & 1+0.052 \\ 
\hline 
\end{tabular} 
}
\caption{The HOD parameters corresponding to $s = \pm 0.2$ that minimizes the $\chi^2(d_{\perp}w)$. The $\chi^2$ values are listed in the second column of Table~\ref{tab:chi2_galprof}.}
\label{tab:besthod_galprof}
\end{table}

\begin{table}
\centering
\scalebox{1}{
\begin{tabular}{ c | c c c }
\hhline {====}
$s = 0.2$ & $\chi^2(d_\perp w)$ & $\chi^2(\xi)$ & $\chi^2(\qeff)$  \\ 
\hline
 unsuppressed HOD & 697 & 1008 & 16.7 \\ 

$d_{\perp}w$ suppressed HOD & 17.6 &1229 & 45.9 \\ 
\hline 
\end{tabular} 
}
\caption{The $\chi^2$ values of the projected 2PCF, the anisotropic 2PCF, and the squeezed 3PCF for the $s = 0.2$ case. The second row lists the $\chi^2$ values for the unsuppressed HOD, whereas the third row lists the $\chi^2$ values for the $d_{\perp}w$ suppressed HOD. The HOD parameters for the suppressed HOD are found in the second row of Table~\ref{tab:besthod_galprof}.}
\label{tab:chi2_galprof} 
\end{table}

Table~\ref{tab:chi2_galprof} lists the $\chi^2$ in the projected 2PCF, the anisotropic 2PCF, and the squeezed 3PCF for the unsuppressed and suppressed HODs. We see that for the $s = 0.2$ case, the suppressed HOD has a residual $\chi^2(d_{\perp}w) \approx 17.6$, down from $\chi^2(d_{\perp}w) \approx 697$ for the unsuppressed HOD. Meanwhile, the $\chi^2(\xi)$ has actually increased from 1008 to 1229. This means that while we have significantly suppressed the projected 2PCF signal, there is still a strong signal residual in the anisotropic 2PCF. Moreover, the signal in the $\chi^2$ in the squeezed 3PCF has actually increased significantly from 16.7 to 45.9, yielding a stronger signal as a result of the suppression. 

\begin{figure*}
\centering
 \hspace*{-0.4cm}
\includegraphics[width=7in]{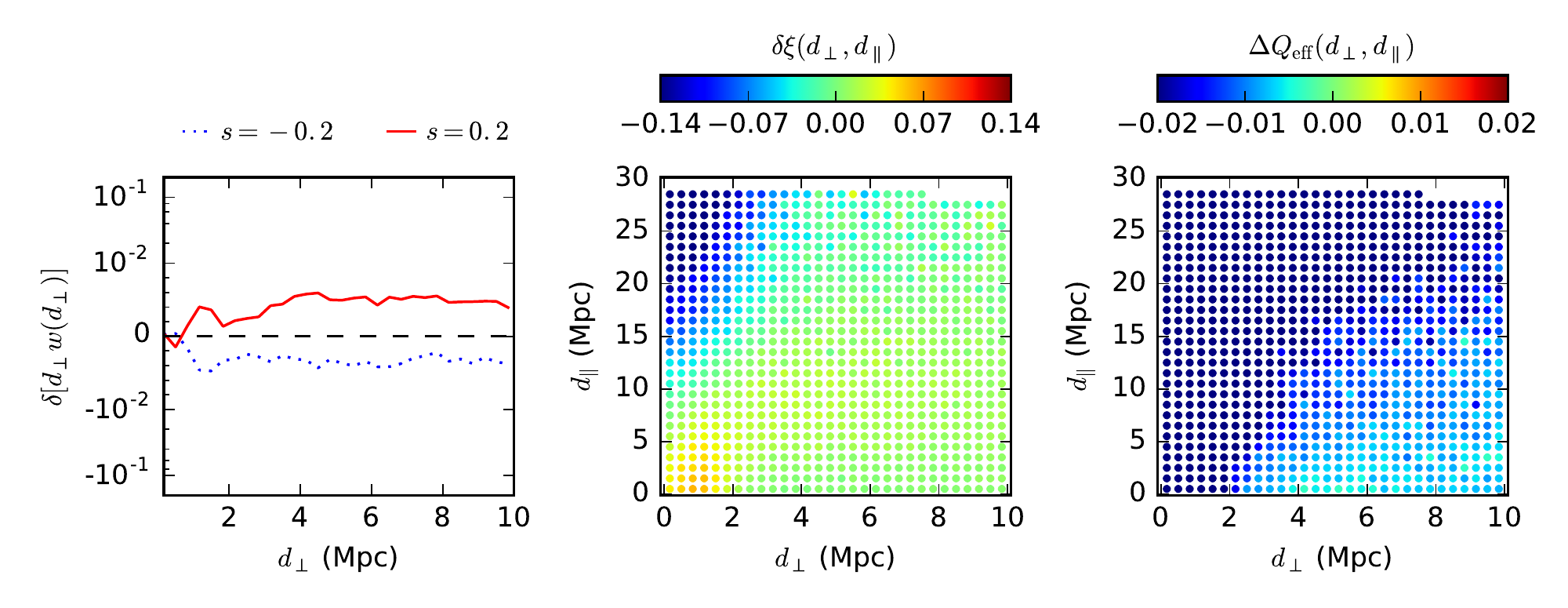}
\vspace{-0.6cm}
\caption{The perturbations to the projected 2PCF, anisotropic 2PCF, and the squeezed 3PCF for the suppressed HODs given in Table~\ref{tab:besthod_galprof}. Left panel: the fractional perturbation to the projected 2PCF for $s = 0.2$ (solid red) and $s = -0.2$ (dotted blue). The y-axis is in linear scale from $-10^{-2}$ to $10^{-2}$ and in log scale beyond $10^{-2}$. Note how the perturbations are now significantly smaller than that shown in the left panel of Figure~\ref{fig:triple_s}. Middle panel: the relative perturbation to the anisotropic 2PCF, where $\delta\xi = [\xi'(s = 0.2)-\xi'(s=-0.2)]/\xi(s=0)$. Right panel: the absolute perturbation to the squeezed 3PCF, where $\Delta \qeff = \qeff'(s = 0.2)-\qeff'(s=-0.2)$. We use $'$ to indicate that these correspond to the suppressed HODs.}
\label{fig:triple_s_suppressed}
\end{figure*}

Figure~\ref{fig:triple_s_suppressed} shows the perturbation to the projected 2PCF, anisotropic 2PCF, and the squeezed 3PCF due to the suppressed HODs given in Table~\ref{tab:besthod_galprof}.
The left panel shows the perturbation to the projected 2PCF for the $s = 0.2$ case (solid red) and the $s = -0.2$ case (dotted blue) for the suppressed HODs.
Comparing to the left panel of Figure~\ref{fig:triple_s}, we see that the difference between the blue and red curves are reduced from $\sim 8\%$ to $\leq 1\%$, consistent with the largely suppressed $\chi^2$ values.
We have thus identified two different HOD models that are nearly degenerate in the projected 2PCF. It is worth pointing out that since the uncertainty in the emulator is $<0.4\%$, the maximum suppression of the projected 2PCF using the emulator is fundamentally limited by the uncertainties in the emulator. Thus, it is possible that there exists a stronger suppression that our emulator has not yet identified. This is an important caveat to keep in mind for the rest of this paper. 

The middle panel and the right panel show the difference between the two suppressed HODs in the anisotropic 2PCF and the squeezed 3PCF. Visually, it is clear that the degeneracy in the projected 2PCF is broken in both the anisotropic 2PCF and the squeezed 3PCF, consistent with the large $\chi^2$ values we have for these two statistics even after suppression.  Visually comparing the middle panel with the left panel, we see that the small signal in the projected 2PCF is effectively a combination of an increase in the anisotropic 2PCF at small $d_{\parallel}$ and a decrease at large $d_{\parallel}$. 

To understand this phenomenon, we refer back to Table~\ref{tab:besthod_galprof}. Consider the $s = 0.2$ case for example. On the one hand, a positive $s$ parameter moves galaxies towards the halo outskirts, boosting the number of large-separation pairs and decreasing the number of very close pairs. On the other hand, the decrease in $\alpha$ and $M_1$ moves galaxies from more massive halos into less massive halos, which produces fewer large-separation pairs and increases the number of small-separation pairs. This counteracts the effects due to satellite distribution modulation. By moving galaxies into less massive halos, we also effectively decrease the velocity dispersion of the pairs along the LOS, thus leading to less RSD on average and a significant decrease in the anisotropic 2PCF towards high $d_{\parallel}$. 

As shown in the right panel of Figure~\ref{fig:triple_s_suppressed}, the difference between the suppressed HODs in the squeezed 3PCF is characterized by a significant drop in the average pair bias. This again can be explained by the fact that the $\chi^2(d_\perp w)$ minimization decreases $\alpha$ and $M_1$ to counteract the increase in $s$, thus moving galaxies to less massive halos. This reduces the average pair bias and leads to the decrease in the squeezed 3PCF. 

To summarize, we have identified a near degeneracy in the projected 2PCF involving the $s$ parameter of the generalized HOD. Both the $\chi^2$ values and the visualizations have shown that this degeneracy is broken by the anisotropic 2PCF and the squeezed 3PCF, thus confirming the extra constraining power that the anisotropic 2PCF and the squeezed 3PCF have on the generalized HOD. 

\subsection{Suppressing the anisotropic 2PCF signal with velocity bias}
\label{subsec:discussion_ani}

We have shown in the last subsection that by suppressing the signal in the projected 2PCF, we are left with a significant signal in the anisotropic 2PCF. 
In this section, we aim to identify two generalized HOD prescriptions that are nearly degenerate in the anisotropic 2PCF, which would reveal the extra constraining power of the squeezed 3PCF on HOD models. 

Notice that in the middle panel of Figure~\ref{fig:triple_s_suppressed}, the structure of the $\xi$ signal is largely along the LOS. This suggests that the velocity bias parameter, which produces a similar signal in $\xi$ (Figure~\ref{fig:triple_sv}), could significantly suppress this signal. However, we will only incorporate the satellite velocity bias parameter $s_v$ because our tests show that the central velocity bias parameter produces negligible $\xi$ signal. 

We also omit the closest approach parameter $s_p$ because it essentially behaves as a combination of satellite distribution generalization and velocity bias generalization, and its signatures on the 2PCF and squeezed 3PCF are similar to that of satellite distribution parameter $s$ (Figure~\ref{fig:triple_sp}). The inclusion of the assembly bias parameter is deferred to the next subsection. 

\begin{figure*}
\centering
 \hspace*{-0.4cm}
\includegraphics[width=7in]{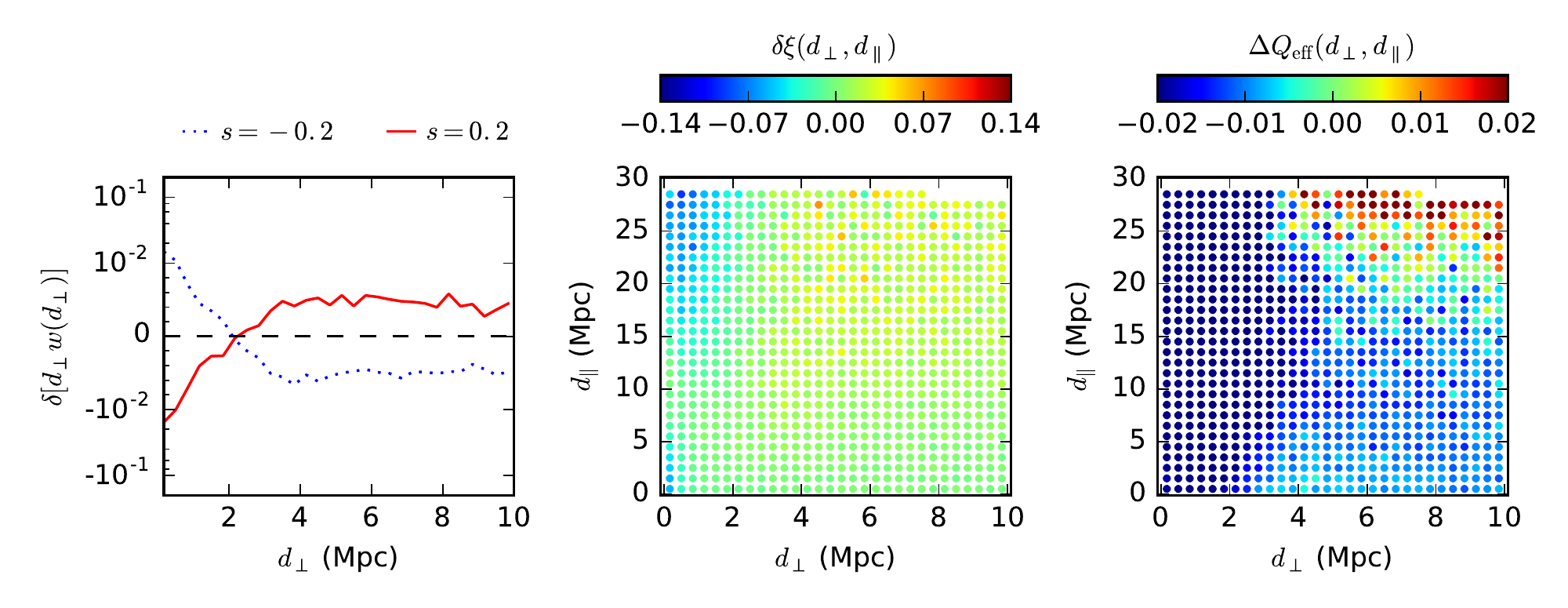}
\vspace{-0.6cm}
\caption{The perturbations to the projected 2PCF, anisotropic 2PCF, and the squeezed 3PCF for the suppressed HODs given in Table~\ref{tab:besthod_s_sv}. Left panel: the fractional perturbation to the projected 2PCF for $s = 0.2$ (solid red) and $s = -0.2$ (dotted blue). The y-axis is in linear scale from $-10^{-2}$ to $10^{-2}$ and in log scale beyond $10^{-2}$. Middle panel: the relative perturbation to the anisotropic 2PCF between the $s = 0.2$ and $s = -0.2$ cases, where $\delta\xi = [\xi'(s = 0.2)-\xi'(s=-0.2)]/\xi(s=0)$. Right panel: the absolute perturbation to the squeezed 3PCF, where $\Delta \qeff = \qeff'(s = 0.2)-\qeff'(s=-0.2)$. We use $'$ to indicate that these correspond to the suppressed HODs.}
\label{fig:triple_s_sv_suppressed}
\end{figure*}

\begin{table*}
\centering
\scalebox{1}{
\begin{tabular}{ c | c c c c c }
\hhline {======}
$s$ & $\log_{10} M_{\rm{cut}}$ & $ \log_{10} M_1$ & $ \sigma$  & $\alpha$ & $s_v$ \\ 
\hline
0.2 & 13.35--0.013 & 13.8--0.0074 & 0.85--0.050 & 1--0.023 & 0.097 \\ 

--0.2 & 13.35+0.011 & 13.8+0.0076 & 0.85+0.050 & 1+0.026 & --0.105 \\ 
\hline 
\end{tabular} 
}
\caption{The HOD parameters corresponding to $s = \pm 0.2$ that minimizes $\chi^2(\xi)$. For the $s = 0.2$ case, the $\chi^2(\xi)$ of the unsuppressed HOD against the baseline HOD is 1008, whereas the $\chi^2(\xi)$ of the suppressed HOD is 265.}
\label{tab:besthod_s_sv}
\end{table*}

\begin{table}
\centering
\scalebox{1}{
\begin{tabular}{ c | c c c }
\hhline {====}
$s = 0.2$ & $\chi^2(d_\perp w)$ & $\chi^2(\xi)$ & $\chi^2(\qeff)$  \\ 
\hline
 unsuppressed HOD & 697 & 1008 & 16.7 \\ 

$\xi$ suppressed HOD & 110 & 265 & 41.2 \\ 
\hline 
\end{tabular} 
}
\caption{The $\chi^2$ values of the projected 2PCF, the anisotropic 2PCF, and the squeezed 3PCF for the $s = 0.2$ case. The first row lists the $\chi^2$ values for the unsuppressed HOD, whereas the second row lists the $\chi^2$ values for the $\xi$ suppressed HOD.}
\label{tab:chi2_s_sv} 
\end{table}

We follow the same procedure as in the the previous subsection. We construct an anisotropic 2PCF emulator with $n=6$ input HOD parameters: four baseline parameters $\log_{10} M_{\rm{cut}}, \log_{10} M_1, \sigma, \alpha$, and 2 generalization parameters $s, s_v$. 
In order to suppress the $\xi$ signal, we minimize the $\chi^2(\xi)$, which is computed from the difference in the anisotropic 2PCF from that of the baseline HOD. 
We minimize $\chi^2(\xi)$ in a similar fashion as for the projected 2PCF. We start with two unsuppressed HODs, one with $s = 0.2$ and the other with $s = -0.2$, and the other 5 parameters start at their Z09 values. Then we vary the other 5 parameters and identify parameter prescriptions that minimize $\chi^2(\xi)$. 
The best-case HOD prescriptions that minimize $\chi^2(\xi)$ are given in Table~\ref{tab:besthod_s_sv} whereas the $\chi^2$ values for the $s = 0.2$ case are displayed in Table~\ref{tab:chi2_s_sv}. 

The second row of Table~\ref{tab:chi2_s_sv} shows the $\chi^2$ values before suppression for the $s = 0.2$ case, whereas the third row shows the $\chi^2$ values after suppression. We see that $\chi^2(\xi)$ decreased from 1008 down to 265, whereas the $\chi^2(d_\perp w)$ only decreased down to 110, much larger than the residual $\chi^2(d_\perp w) = 17.6$ that we obtained by suppressing the projected 2PCF signal. This suggests that while we do manage to significantly suppress the anisotropic 2PCF, we have to sacrifice some of the degeneracy in the projected 2PCF. The $\chi^2(\qeff)$ increased from 16.7 to 41.2, again revealing a stronger squeezed 3PCF signal after suppression. 

Figure~\ref{fig:triple_s_sv_suppressed} shows the perturbation to the projected 2PCF, anisotropic 2PCF, and the squeezed 3PCF for the suppressed HODs given in Table~\ref{tab:besthod_s_sv}. Comparing the middle panel to the middle panel of Figure~\ref{fig:triple_s}, we see that the introduction of velocity bias parameters does significantly suppress the change in the anisotropic 2PCF. However, there is still a $\sim 5\%$ difference in some regions of the anisotropic 2PCF. The left panel shows that these two suppressed HODs are $\leq 2\%$ different in the projected 2PCF. The right panel also shows that the suppressed HODs have a $> 2\%$ signal in the squeezed 3PCF towards small $d_{\perp}$. The large perturbations towards the top right corner of the plot are dominated by shot-noise. 

This shows that the introduction of satellite velocity bias of our chosen functional form does allow us to significantly suppress the signal in the anisotropic 2PCF but it does not do a proper job maintaining a near degeneracy in the projected 2PCF. Comparing the final $\chi^2$ values, we see that we have yet to generate a stronger signal in the squeezed 3PCF than in the anisotropic 2PCF, underscoring the constraining power of the anisotropic 2PCF on galaxy-halo connection models. 

\subsection{Suppressing the anisotropic 2PCF signal with assembly bias}
\label{subsec:discussion_A}

\begin{figure*}
\centering
 \hspace*{-0.4cm}
\includegraphics[width=7in]{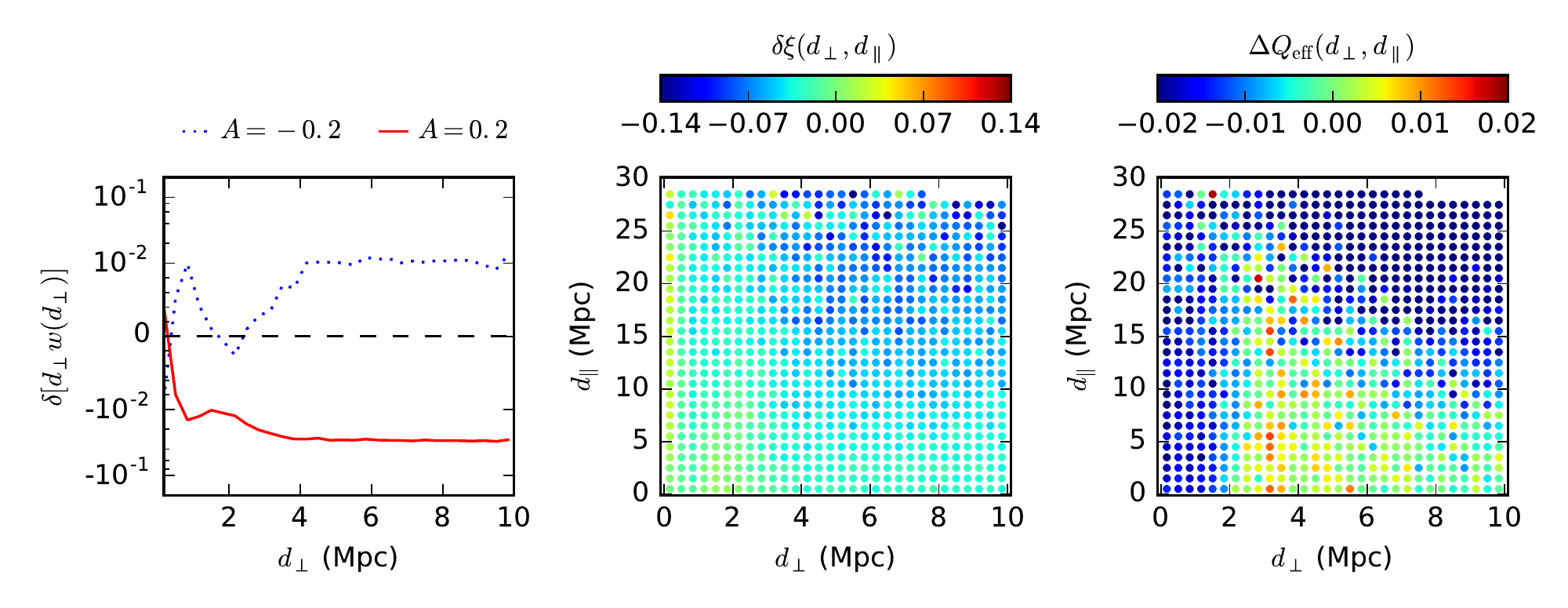}
\vspace{-0.6cm}
\caption{The perturbations to the projected 2PCF, anisotropic 2PCF, and the squeezed 3PCF for the suppressed HODs given in Table~\ref{tab:besthod_s_sv_A}. Left panel: the fractional perturbation to the projected 2PCF for $A = 0.2$ (solid red) and $A = -0.2$ (dotted blue). The y-axis is in linear scale from $-10^{-2}$ to $10^{-2}$ and in log scale beyond $10^{-2}$. Middle panel: the relative perturbation to the anisotropic 2PCF between the $A = 0.2$ and $A = -0.2$ cases, where $\delta\xi = [\xi'(A = 0.2)-\xi'(A=-0.2)]/\xi(A=0)$. Right panel: the absolute perturbation to the squeezed 3PCF, where $\Delta \qeff = \qeff'(A = 0.2)-\qeff'(A=-0.2)$. We use $'$ to indicate that these correspond to the suppressed HODs.}
\label{fig:triple_s_sv_A_suppressed}
\end{figure*}

\begin{table*}
\centering
\scalebox{1}{
\begin{tabular}{ c | c c c c c c }
\hhline {=======}
$A$ & $\log_{10} M_{\rm{cut}}$ & $ \log_{10} M_1$ & $ \sigma$  & $\alpha$ & $s$ & $s_v$ \\ 
\hline
0.2   & 13.35--0.0018 & 13.8+0.025 & 0.85--0.042 & 1+0.018 & 0.153 &  0.013 \\ 

--0.2 & 13.35--0.0056 & 13.8+0.001 & 0.85+0.004 & 1+0.007 & --0.061 &  0.059 \\ 
\hline 
\end{tabular} 
}
\caption{The HOD parameters corresponding to $A = \pm 0.2$ that minimizes $\chi^2(\xi)$. For the $A = 0.2$ case, the $\chi^2(\xi)$ of the unsuppressed HOD against the baseline HOD is 1266, whereas the $\chi^2(\xi)$ of the suppressed HOD is 378.}
\label{tab:besthod_s_sv_A}
\end{table*}

\begin{table}
\centering
\scalebox{1}{
\begin{tabular}{ c | c c c }
\hhline {====}
$A = 0.2$ & $\chi^2(d_\perp w)$ & $\chi^2(\xi)$ & $\chi^2(\qeff)$  \\ 
\hline
 unsuppressed HOD & 826 & 1266 & 97.4 \\ 

$\xi$ suppressed HOD & 272 & 378 & 102 \\ 
\hline 
\end{tabular} 
}
\caption{The $\chi^2$ values of the projected 2PCF, the anisotropic 2PCF, and the squeezed 3PCF for the $A = 0.2$ case. The first row lists the $\chi^2$ values for the unsuppressed HOD, whereas the second row lists the $\chi^2$ values for the $\xi$ suppressed HOD.}
\label{tab:chi2_s_sv_A} 
\end{table}

As we discussed in Section~\ref{sec:signatures}, the squeezed 3PCF contains new information on the average halo mass of the pairs. While the satellite distribution generalization and the velocity bias generalizations does not directly modulate the average halo mass of the pairs, the assembly bias generalization modulates the halo mass of the pairs by effectively moving pairs into halos of different mass. Thus, we explore the possibility that the assembly bias generalization, combined with satellite distribution generalization and velocity bias generalization, can bring out the extra constraining power of the squeezed 3PCF. 

We build an emulator of the anisotropic 2PCF that takes the 4 baseline HOD parameters and 3 generalization parameters $s$, $s_v$, and $A$ as inputs ($n=7$). 
We start with two unsuppressed HODs, one with $A = 0.2$ and the other with $A = -0.2$. The other 6 parameters start at their Z09 values. We perturb the 6 parameters and find prescriptions that minimize the $\chi^2(\xi)$ function. The best-case HODs are given in Table~\ref{tab:besthod_s_sv_A} and the $\chi^2$ values for the $A = 0.2$ case are displayed in Table~\ref{tab:chi2_s_sv_A}. The 2PCF and 3PCF signatures of these two best-case HODs are shown in Figure~\ref{fig:triple_s_sv_A_suppressed}.

Table~\ref{tab:chi2_s_sv_A} shows that the suppression of the $\xi$ signal has significantly reduced the $\chi^2(\xi)$ from 1266 down to 378. However, the large residual $\chi^2$ in the projected 2PCF $\chi^2(d_\perp w) = 272$ and the anisotropic 2PCF suggest that we are still left with a strong signal in our 2-point statistics. The $\chi^2$ on the squeezed 3PCF has increased slightly from 97.4 to 102, again showing a different response to HOD parameter changes compared to the 2PCF. 

Visually comparing the middle panel of Figure~\ref{fig:triple_A} and the middle panel of Figure~\ref{fig:triple_s_sv_A_suppressed}, we see that the inclusion of satellite distribution parameter and velocity bias parameter does enable us to significantly suppress the anisotropic 2PCF signal due to assembly bias from $\leq 10\%$ to $\leq 4\%$. The left panel of Figure~\ref{fig:triple_s_sv_A_suppressed} shows that these two best-case HODs are different by $\leq 4\%$ in the projected 2PCF, not significantly lower compared to the unsuppressed signal shown in the left panel of Figure~\ref{fig:triple_A}. The right panel shows the difference between these two HODs in the squeezed 3PCF, which appears similar to the unsuppressed signal shown in the right panel of Figure~\ref{fig:triple_A}. 

Both the $\chi^2$ values and the visualizations suggest that while we have significantly suppressed the anisotropic 2PCF signal with the inclusion of velocity bias and assembly bias, we have not yet identified two generalized HODs with a stronger signal in the squeezed 3PCF than in the anisotropic 2PCF, and that our current generalizations have not revealed regions of the generalized HOD parameter space that the squeezed 3PCF is uniquely sensitive to.

However, this does not mean that the squeezed 3PCF has the same constraining power on the generalized HOD as the anisotropic 2PCF. These two statistics offer fundamentally different information on clustering. The anisotropic 2PCF contains information on pair counts whereas the squeezed 3PCF offers information on pair bias.

It is also important to point out that the squeezed 3PCF by itself is still an informative test of the HOD framework. 
We find that no matter how we adjust the generalized HOD parameters to suppress the 2PCF signal, the $\chi^2$ values for the anisotropic 2PCF are consistently higher than that of the squeezed 3PCF. This suggests that for all of the parameter choices we have tested so far, the anisotropic 2PCF yields a much stronger signal than the squeezed 3PCF, which in turn suggests that the 2PCF is predicting the squeezed 3PCF to some precision better than the statistical error. Thus, any significant observed deviation from the prediction in the squeezed 3PCF would challenge the validity of the standard HOD parametrizaiton.Thus, the squeezed 3PCF still serves as an important null test of any HOD model.

 A rather interesting trend we see in this section is that when we suppress the signals in the projected 2PCF and the anisotropic 2PCF, the $\chi^2(\qeff)$ actually increases consistently. This suggests that the squeezed 3PCF reacts to HOD changes differently than the projected and anisotropic 2PCF. 
This point can also be seen visually. Comparing the unsuppressed (right panel of Figure~\ref{fig:triple_s}) and suppressed signal (right panel of Figure~\ref{fig:triple_s_sv_suppressed}) in the squeezed 3PCF due to satellite distribution parameter, we see that the magnitude of the squeezed 3PCF signal has dramatically increased with the suppressed HOD.
 
 This suggests that with further generalizations of the HOD model, we can identify regions of the parameter space that produces the strongest signal in the squeezed 3PCF and potentially use the squeezed 3PCF to break model degeneracies in the 2PCFs. One generalization of interest is where $s, s_v, \alpha_c$ parameters are functions of halo mass. Another such generalization is an additional satellite velocity bias term that is not virialized in the halo potential well. This is motivated by considerations of the first infall of subhalos. 
 
In addition to introducing further generalizations, we plan to make several improvements to our routine. One improvement is to increase the accuracy of our 2PCF emulators, potentially by adopting a more flexible analytic model. We also plan on running more cosmological simulation boxes to improve the covariance calculations. 
We defer these discussions to a future paper.

\section{Conclusions}
\label{sec:conclusions}

In this paper, we introduce a generalized and differentiable HOD routine (GRAND-HOD) that incorporates generalizations in satellite distribution, velocity bias, closest approach radius, and assembly bias. Utilizing mocks generated from the \textsc{ABACUS} cosmological simulations, we showcase the signature of these generalizations in the projected 2PCF, the anisotropic 2PCF, and the squeezed 3PCF. To test the extra constraining power of the squeezed 3PCF on the generalized HOD, we search for generalized HOD prescriptions that are degenerate in the 2PCF but distinguishable in the squeezed 3PCF. We are able to identify near degeneracies in the projected 2PCF with a mixture of the baseline HOD parameters and the satellite distribution parameter. Such degeneracy is broken by both the anisotropic 2PCF and the squeezed 3PCF. We have yet to identify a degeneracy in the anisotropic 2PCF that results in a smaller $\chi^2$ difference than the accompanying $\chi^2$ change in the squeezed 3PCF; this implies that a discordant squeezed 3PCF measurement could falsify the particular HOD model space. Alternatively, it is possible that further generalizations of the HOD model would open opportunities for the squeezed 3PCF to provide novel parameter measurements.  We stress that the squeezed 3PCF is a physically distinct measurement, focused on the mass of regions containing close pairs rather than the simple frequency of those pairs, and thereby provides a key opportunity for models of the galaxy-halo connection.

\section*{Acknowledgements}
We would like to thank Andrew Hearin, Joshua Speagle, Ryuichiro Hada, and Alexie Leauthaud for useful discussions and feedback on the GRAND-HOD package. DJE and LG are supported by National Science Foundation grant AST-1313285. DJE is additionally supported by U.S. Department of Energy grant DE-SC0013718 and as a Simons Foundation Investigator.

The GeneRalized ANd Differentiable HOD (\textsc{GRAND-HOD}) package is publicly available at \url{https://github.com/SandyYuan/GRAND-HOD}.
The \textsc{ABACUS} simulations used in this paper are available at \url{https://lgarrison.github.io/AbacusCosmos}.
 
\bibliographystyle{aasjournal}
\bibliography{biblio}
\end{document}